\documentclass[reprint, superscriptaddress, aps, pre]{revtex4-2}

\usepackage{amsmath}
\usepackage{graphicx}
\usepackage{epstopdf}
\usepackage{color}
\usepackage{amssymb}
\usepackage{amsmath}
\usepackage{tabularx}
\usepackage{bm}
\usepackage{hyperref}
\usepackage{ltxtable}
\usepackage{natbib}
\usepackage{longtable}
\usepackage[normalem]{ulem}
\usepackage{float}

\newcommand{\la}{\left\langle}
\newcommand{\ra}{\right\rangle}
\newcommand{\be}{\begin{equation}}
	\newcommand{\ee}{\end{equation}}
\newcommand{\bse}{\begin{subequations}}
	\newcommand{\ese}{\end{subequations}}
\newcommand{\bea}{\begin{eqnarray}}
	\newcommand{\eea}{\end{eqnarray}}
\newcommand{\ba}{\begin{array}}
	\newcommand{\ea}{\end{array}}
\newcommand{\balign}{\begin{align}}
		\newcommand{\ealign}{\end{align}}

\usepackage[usenames,dvipsnames]{xcolor}
\hypersetup{
	colorlinks,
	citecolor=blue,
	filecolor=blue,
	linkcolor=blue,
	urlcolor=blue}

\begin{document}
	
	\title{Critical dimension for  hydrodynamic turbulence}
	\author{Mahendra K. Verma}
	\email{mkv@iitk.ac.in}
	\affiliation{Department of Physics, Indian Institute of Technology Kanpur, Kanpur 208016, India}
	\date{\today}

\begin{abstract}

Hydrodynamic turbulence exhibits nonequilibrium behaviour with $k^{-5/3}$ energy spectrum, and equilibrium behaviour with $k^{d-1}$ energy spectrum and zero viscosity, where $d$ is the space dimension. Using recursive renormalization group {in Craya-Herring basis}, we show that the nonequilibrium solution is valid only for $d < 6$, whereas equilibrium solution with zero viscosity is the only solution for $d>6$. Thus, $d=6$ is the critical dimension for hydrodynamic turbulence. In addition, we show that the energy flux changes sign from positive to negative near $d=2.15$.  We also compute the energy flux and Kolmogorov's constants for various $d$'s, and observe that our results are in good agreement with past numerical results.
\end{abstract}


\maketitle

\section{Introduction}

Field theoretic tools help explain complex phenomena in high-energy physics, condensed-matter physics, statistical physics, and turbulence~\citep{Peskin:book:QFT,Lancaster:book,Zinn-Justin:book,Vasilev:book:RG_critical}. For example, Wilson and coworkers~\cite{Wilson:PR1974} constructed a theory for the second-order phase transition that goes beyond the mean field theory of Landau~\cite{Ma:book:StatMech}. In Wilson's theory, the nonlinear term yields nontrivial scaling for $d<4$, but it become irrelevant for $d \ge 4$. Therefore, the critical dimension for the second-order phase transition is 4. In this paper, we compute the critical dimension for hydrodynamic turbulence.

The frameworks of quantum field theory and statistical field theory have been extended to hydrodynamic turbulence. Prominent field-theoretic computations for hydrodynamic turbulence are {Direct Interaction Approximation} (DIA)  \cite{Kraichnan:JFM1959}, { Renormalization Group (RG)~\cite{Forster:PRA1977, Yakhot:JSC1986,Adzhemyan:book:RG}, Generating Functionals \cite{DeDominicis:PRA1979,Adzhemyan:book:RG},} Martin-Siggia-Rose (MSR)  formalism~\cite{Martin:PRA1973},  Recursive Renormalization Group \cite{McComb:PRA1983,McComb:book:HIT,Zhou:PRA1988}, Functional Renormalization~\cite{Canet:JFM2022}. Other field theory works on  hydrodynamic turbulence are \cite{Eyink:PRE1993,Moriconi:PRE2004,Canet:JFM2022,Arad:PRE1999,Biferale:PR2005,Bos:JFM2013,Kaneda:FDR2007}. These works are reviewed in   \citet{Orszag:CP1973} and \citet{Zhou:PR2010}. Most of the prominent field theory works are for three dimensions (3D), where the RG analysis predicts that the energy spectrum $E(k) \propto k^{-5/3}$, and that the renormalization viscosity $\nu(k)$ scales as $k^{-4/3}$ with the renormalization constant around 0.40.  Some calculations (e.g., \cite{Yakhot:JSC1986}) employ particular forcing, whereas some others employ self-consistent procedure \cite{McComb:PRA1983,Zhou:PR2010}. In comparison, RG works on two-dimensional (2D) hydrodynamic turbulence is limited. In one such works,  \citet{Olla:PRL1991}  obtained two different spectral regimes: $k^{-3}$ energy spectrum with a  constant enstrophy flux at large wavenumbers, and $k^{-5/3}$ spectrum with a constant   energy flux at small wavenumbers.  For the $k^{-5/3}$ spectral regime, \citet{Olla:PRL1991} derived the renormalization constant to be 0.642 and the Kolmogorov constant to be 6.45. \citet{Nandy:IJMPB1995}  employed self-consistent mode-coupling scheme and obtained similar constants.

The energy transfers and fluxes of hydrodynamic turbulence are also computed using field theory. \citet{Kraichnan:JFM1959} employed direct interaction approximation (DIA) for these computations. Later, eddy-damped quasi-normal Markovian approximation (EDQNM) and other schemes have been used for the flux calculations~\cite{Orszag:CP1973}. \citet{Verma:PR2004,Verma:book:ET}  computed the energy fluxes using the mode-to-mode energy transfers.  The equation for the energy flux yields Kolmogorov's constant~\cite{Orszag:CP1973}.

\citet{Fournier:PRA1978} employed EDQNM procedure to compute the stable energy spectra for various space dimensions, denoted by $d$. They showed that the energy flux changes sign from positive to negative near   $d = 2.05$ as $d$ decreases from 3 to 2.  \citet{Gotoh:PRE2007} employed Lagrangian Renormalized Approximation and showed that the energy transfer in 4D is more efficient compared to that in 3D.  Consequently, the Kolmogorov's constant for 4D, $K_\mathrm{Ko} = 1.31$, is smaller than that for 3D,  $K_\mathrm{Ko} = 1.72$. \citet{Gotoh:PRE2007} verified the field-theoretic predictions using numerical simulations. \citet{Berera:PF2020} observed similar results in their numerical simulations, for example, $K_\mathrm{Ko} = 1.7$ and 1.3 for 3D and 4D respectively.

{In statistical and quantum field theory, the parameters of theory (e.g., coupling constant and mass) depend critically on the space dimension~\citep{Peskin:book:QFT,Lancaster:book,Zinn-Justin:book,Vasilev:book:RG_critical}. For example, the fluctuations in $\phi^4$ theory obey Gaussian property for $d \ge 4$. Hence, $d=4$ is called the upper critical dimension for the $\phi^4$ theory. Theorists have been exploring whether such a upper critical dimension exists for hydrodynamic turbulence. For example, \citet{Adzhemyan:JPA2008} showed that the  Kolmogorov constant $K_\mathrm{Ko} \propto d^{1/3}$, which leads to  the energy flux $\epsilon_u \propto K_\mathrm{Ko}^{-3/2} \propto d^{-1/2} \rightarrow 0$ as  $d \rightarrow \infty$. Similarly, \citet{Fournier:JPA1978} showed that intermittency, which is a reflection of nongaussian nature of the fluctuations, vanishes as $d \rightarrow \infty$. These observations indicate that the velocity fluctuations possibly exhibit Gaussian behaviour for large $d$. In this paper, we explore this issue for hydrodynamic turbulence using RG calculation in Craya-Herring basis that provides detailed picture of interactions in hydrodynamic turbulence. }

In this paper, we compute the renormalized viscosity using recursive renormalization,  and Kolmogorov's constant using energy transfers for various dimensions.  For these computations, we 
employ Craya-Herring basis ~\citep{Craya:thesis,Herring:PF1974,Sagaut:book,Verma:book:ET} that simplifies the field-theoretic calculations of turbulent flows  dramatically.  In addition, this basis allows separate computations of the renormalized viscosities and energy transfers for each component, thus yielding finer details of turbulence without complex tensor algebra.    In addition to the above simplification, we deviate from the conventional $\int dpdq \delta({\bf k-p-q})$ integrals to $\int dp d\gamma$, where $\gamma$ is the angle between \textbf{k} and \textbf{p} in a triad $(k,p,q)$. This new scheme simplifies the asymptotic analysis, as well as the evaluation of the singular integrals of energy fluxes~\cite{Verma:arxiv2023}.

Using the above techniques, we compute the renormalized viscosity and Kolmogorov's constant for various $d$. We show that $\nu(k) \sim k^{-4/3}$ solution exists only for $d < 6$, whereas  $\nu = 0$ is the solution of the RG equation beyond $d = 6$. Hence, the critical dimension for hydrodynamic turbulence is 6.  In addition, we also compute the energy flux in the inertial range that yields Kolmogorov's constant.  Our Kolmogorov constants for various dimensions are in good agreement with the past works~\cite{Gotoh:PRE2007,	Berera:PF2020}.

The outline of the paper is as follows: In Sections 2, we introduce the relevant hydrodynamic equations in Craya-Herring basis. In Section 3, we describe the renormalization group analysis for hydrodynamic turbulence using Craya-Herring basis. Section 4  contains discussions on the energy transfers in a triad, as well as the energy fluxes for various $d$.  Section 5 provides a brief discussion on the fractional energy transfers. Section 6 reproduces the RG and energy flux computations for Kraichnan's $k^{-3/2}$ energy spectrum.   We conclude in Section 7.

\section{Governing Equations and Framework}

In Fourier space, the equations for the incompressible Navier-Stokes equations in $d$ dimensions  are~\citep{Lesieur:book:Turbulence,Verma:book:ET}
\bea
(\partial_t +\nu k^2){\bf u} (\mathbf{k},t) 
& = & {- i \int \frac{d\bf p}{(2\pi)^d}  }  \{ {\bf k} \cdot  {\bf u}({\bf q}, t) \} {\bf u}({\bf p}, t)   \nonumber \\
&&  -i {\bf k} p (\mathbf{k},t)   + {\bf F}_u({\bf k},t) , \label{eq:uk}\\
{\bf k\cdot u} (\mathbf{k},t) & = & 0, \label{eq:k_uk_zero}
\eea 
where ${\bf k = p+q}$;  { ${\bf u}, p$ are the velocity and pressure fields respectively; $\nu$ is the kinematic viscosity; and ${\bf F}_u$ is the external forcing, which is active at large scales, as in Kolmogorov's theory of turbulence.}  The transformation from real space to Fourier space and vice versa are as follows~\citep{Peskin:book:QFT}:
\bea
{\bf u}({\bf r},t) & = & \int \frac{d\bf k}{(2\pi)^d} {\bf u}({\bf k}, t) \exp( i {\bf k \cdot r}), 
\label{eq:Fourier_ktox} \\
{\bf u}({\bf k}, t) & = &   \int d{\bf r} [{\bf u}({\bf r},t)   \exp( -i {\bf k \cdot r})];
\label{eq:Fourier_xtok}
\eea
and the pressure field is determined using the following equation:
\be 
p(\mathbf{k},t) =  - \frac{i}{k^2} {\bf k} \cdot {\bf F}_u({\bf k},t)  - \frac{1}{k^2}  \int \frac{d\bf p}{(2\pi)^d}    \{ {\bf k} \cdot  {\bf u}({\bf q}, t) \}   \{ {\bf k} \cdot  {\bf u}({\bf p}, t) \} \}  
\label{eq:Pk}
\ee
with ${\bf k = p+q}$.

The equation for the \textit{modal energy}   $E({\bf k}) = |{\bf u(k)}|^2/2$ is~\citep{Dar:PD2001,Verma:PR2004}
\bea
(\partial_t +2 \nu k^2) E(\mathbf{k},t) & = &   \int \frac{d\bf p}{(2\pi)^d}    S^{uu}({\bf k|p|q})  \nonumber \\
&&  + \Re[ {\bf F}_u({\bf k},t) \cdot {\bf u}^*({\bf k}, t) ]  ,
\label{eq:Euk}
\eea
where
\be
S^{uu}({\bf k|p|q}) = \Im \left[   \{ {\bf k} \cdot  {\bf u}({\bf q}, t) \}   \{ {\bf u}({\bf p}, t) \cdot  {\bf u}^*({\bf k}, t) \} \}    \right]
\label{eq:Suu}
\ee
is the \textit{mode-to-mode energy transfer rate} from the \textit{giver} mode ${\bf u(p)}$ to the \textit{receiver} mode ${\bf u(k)}$ with the mediation of mode ${\bf u(q)}$. Here, $\Im$ stands for the imaginary part of the argument.   The energy flux $ \Pi(R)$ is the net nonlinear energy transfer rate from all the modes residing inside the sphere of radius $R$  to the modes outside the sphere. Hence, the  ensemble average of  $ \Pi(R)$ is~\citep{Dar:PD2001,Verma:PR2004,Verma:book:ET}
\be
\la  \Pi(R)  \ra = \int_{R}^\infty \frac{d{\bf k'}}{(2\pi)^d} \int_0^{R} \frac{d{\bf p}}{(2\pi)^d}  \la S^{u u}({\bf k'|p|q}) \ra.
\label{eq:fluid_flux}
\ee

In this paper, we will compute the renormalized viscosity, as well as $ \la S^{u u}({\bf k'|p|q}) \ra$ and  $\la \Pi(k_0) \ra$, using field theory in Craya-Herring basis. In this basis, the basis vectors in 3D are~\citep{Craya:thesis,Herring:PF1974,Sagaut:book}:
\bea
\hat{e}_0({\bf k}) = \hat{k};~~
\hat{e}_1({\bf k}) = \frac{ \hat{k} \times \hat{n}}{|\hat{k} \times \hat{n}|};~~~
\hat{e}_2({\bf k})= \hat{e}_0({\bf k})  \times \hat{e}_1({\bf k}) ,   \label{eq:CH_basis_defn}
\eea 
where  the unit vector $\hat{k}$ is along the wavenumber ${\bf k}$, and the unit vector $\hat{n}$ is chosen along any direction.  For space dimension $d$ greater than 3, we choose additional $d-3$ orthogonal unit vectors that are perpendicular  to $\hat{e}_0({\bf k})$, $\hat{e}_1({\bf k})$, and $\hat{e}_2({\bf k})$.
For an incompressible flow, 
\bea
{\bf u}({\bf k},t) & = & \sum_{j=1}^{d-1} u_j({\bf k},t)  \hat{e}_j ({\bf k}). 
\eea

\begin{figure}
	\begin{center}
		\includegraphics[scale = 1]{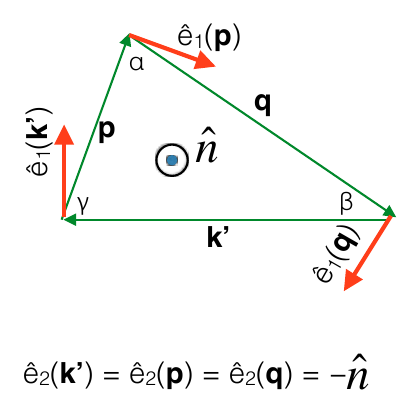}
	\end{center}
	\vspace*{0pt}
	\caption{Craya-Herring basis vectors  for an interacting wavenumber triad $({\bf k',p,q})$.  Reprinted with permission from \citet{Verma:book:ET}.}
	\label{fig:CH_triad}
\end{figure}

In this paper, we will derive the renormalized viscosity and energy flux by summing up contributions from all the interacting triads. Therefore, as a first step, we write down the evolution equations for  $u_j({\bf k},t)$  in a triad.  For the same, we consider a wavenumber triad $({\bf k',p,q})$  with ${\bf k'+p+q} = 0$, and choose $\hat{n} $ as follows~\citep{Waleffe:PF1992,Verma:book:ET}:
\be
\hat{n} = \frac{{\bf q \times p}} {|{\bf q \times p}|}.
\label{eq:hat_n}
\ee
Since ${\bf k = p+q}$, we deduce that ${\bf k' = -k}$.  The Craya-Herring basis vectors for the interacting wavenumbers   are illustrated in Fig.~\ref{fig:CH_triad}.  Note that $ \alpha, \beta$,  $\gamma$ are the angles in front of $ k, p, q $ respectively.  The net nonlinear interaction is a sum over all possible triads. Hence,   the equations for the  ${u}_1$ components of a triad $({\bf k',p,q})$ are~\cite{Verma:book:ET}:
\bea
(\partial_t +\nu k^2){u}_1({\bf k'},t) & = & i k' { \int \frac{d{\bf p}} {(2\pi)^{d}}  } \sin(\beta-\gamma) u_1^*({\bf p},t)  u_1^*({\bf q},t) \nonumber \\
&&  +F_1({\bf k'},t) ,
\label{eq:u1k_dot} \\
(\partial_t +\nu k^2){u}_1({\bf p},t) & = &i p { \int \frac{d{\bf q}} {(2\pi)^{d}}  }  \sin(\gamma-\alpha) u_1^*({\bf q},t) u_1^*({\bf k'},t) \nonumber \\
&&  +F_1({\bf p},t)  , 
\label{eq:u1p_dot}\\
(\partial_t +\nu k^2){u}_1({\bf q},t) & = &i q  {  \int \frac{d{\bf k'}} {(2\pi)^{d}}  } \sin(\alpha - \beta)   u_1^*({\bf p},t) u_1^*({\bf k'},t) \nonumber \\
&& +F_1({\bf q},t) 
\label{eq:u1q_dot}
\eea 
with ${\bf k'+p+q = 0}$, and the angles $ \alpha, \beta$,  $\gamma$ are computed for the respective triads.  The equations for ${u}_2({\bf k'},t)$, ${u}_3({\bf k'},t)$, ..., ${u}_{d-1}({\bf k'},t)$, denoted by ${u}_j({\bf k'},t)$, are similar:
\bea
(\partial_t +\nu k^2){u}_j({\bf k'},t) & = &i k' { \int \frac{d{\bf p}} {(2\pi)^{d}}  }  \{ \sin \gamma u_1^*({\bf p},t)  u_j^*({\bf q},t) \nonumber \\
&& -\sin\beta u_1^*({\bf q},t)  u_j^*({\bf p},t)\} +F_j({\bf k'},t) , \label{eq:u2k_dot}  \nonumber \\ \\
(\partial_t +\nu k^2){u}_j({\bf p},t) & = &i p {\int \frac{d{\bf q}} {(2\pi)^{d}}  }  \{ \sin \alpha u_1^*({\bf q},t)  u_j^*({\bf k'},t) \nonumber \\
&& -\sin\gamma u_1^*({\bf k'},t)  u_j^*({\bf q},t)\} + F_j({\bf p},t) , \label{eq:u2p_dot}  \nonumber \\ \\
(\partial_t +\nu k^2){u}_j({\bf q},t) & = &i q { \int \frac{d{\bf k'}} {(2\pi)^{d}}  }   \{ \sin \beta u_1^*({\bf k'},t)  u_j^*({\bf p},t) \nonumber \\
&& -\sin\alpha u_1^*({\bf p},t)  u_j^*({\bf k'},t)\} +F_j({\bf q},t) 
\label{eq:u2q_dot}  \nonumber \\
\eea 
with ${\bf k'+p+q = 0}$.

The energy flux is  compactly captured by the following mode-to-mode energy transfers in the Craya-Herring basis~\citep{Dar:PD2001,Verma:book:ET}:
\bea
S^{uu}({\bf k'|p|q}) & = &{ \sum_{j=1}^{d-1}}
S^{u_j u_j}({\bf k'|p|q}),
\label{eq:Suu_kpq_CH}
\eea
with
\bea
S^{u_1 u_1}({\bf k'|p|q})& = &k' \sin\beta \cos \gamma  \Im \{u_1({\bf q},t) u_1({\bf p},t) u_1({\bf k'},t) \} ,  \label{eq:Su1u1} \nonumber \\  \\
S^{u_j u_j}({\bf k'|p|q}) & = &- k' \sin\beta \Im \{u_1({\bf q},t) u_j({\bf p},t) u_j({\bf k'},t) \} \label{eq:Su2u2}
\eea
for $ j \in [2..(d-1)]$. An isotropic $d$-dimensional divergence-free flow field has $d-1$ Craya-Herring  components with  
{
\be
\la |u_1({\bf k}|^2 \ra = \la |u_2({\bf k}|^2 \ra = ... = \la |u_{d-1}({\bf k}|^2 \ra   = C({\bf k}) .
\ee
In this paper, we denote $ \la |u_j({\bf k}|^2 \ra = C_j{\bf k})$. } The total kinetic energy is
\bea 
\frac{\la u^2 \ra}{2} & = & \int E(k) dk = \frac{1}{2} \int  \frac{d{\bf k }} {(2\pi)^{d}} (d-1) C({\bf k}) \nonumber \\
& = & \frac{1}{2} \frac{S_d}{(2\pi)^d} (d-1) \int dk  k^{d-1} C({\bf k}), 
\eea
where $E(k)$ is the one-dimensional (1D) shell spectrum, and $S_d = 2 \pi^{d/2}/\Gamma(d/2)$ is the surface area of the $d$-dimensional sphere.  The above equation yields the following  relationship between the modal energy and 1D energy spectrum~\citep{Kraichnan:JFM1959,Leslie:book,Verma:PR2004}:
\be
E(k) = \frac{(d-1)}{2} C({\bf k}) \frac{S_d  k^{d-1} }{ (2\pi)^d}.
\label{eq:Ek_Ck}
\ee

After the above preliminary discussion on the relevant equations, we perform  renormalization group (RG)   and energy transfer analysis for $d$-dimensional hydrodynamic turbulence.

\section{Renormalization Group Analysis of Hydrodynamic Turbulence}
\label{sec:RG}
In this section, we derive the renormalized viscosity using the Craya-Herring basis.   We follow the recursive RG method proposed by McComb, Zhou, and coworkers~\citep{McComb:PRA1983,McComb:book:Turbulence,Zhou:PRA1988}.  Note that the coupling constant, the coefficient in front of the nonlinear term ${\bf u \cdot \nabla u}$, is unchanged under renormalization due to the Galilean invariance~\citep{Forster:PRA1977,McComb:book:HIT}.  Therefore,  vertex renormalization is not required in hydrodynamic turbulence. In addition, in the recursive RG, the forcing or noise is introduced at large scales so as to produce a steady-state with Kolmogorov spectrum.  Hence \textit{noise renormalization} too is avoided in this scheme~\citep{McComb:PRA1983,McComb:book:Turbulence,Zhou:PRA1988}, and the energy spectrum is taken as $k^{-5/3}$. Note that such a choice for $E(k)$ is as  arbitrary as the choice of noise  that yields Kolmogorov's spectrum (as is done in noise renormalization~\cite{Yakhot:JSC1986}).  

The evolution equations for $u_1$ differs from the other components. Therefore, we expect that $u_1$'s renormalized viscosity, denoted by $\nu_1(k)$,  differs from that of others, which is denoted by $\nu_2(k)$. Note that the renormalized viscosities of $u_2, u_3, ..., u_{d-1}$ are the same due to the symmetries of Eqs.~(\ref{eq:u2k_dot}-\ref{eq:u2q_dot}).

\subsection{Renormalization of  $u_1$ Component }
\label{sec:RG_u1} 

\begin{figure}
	\begin{center}
		\includegraphics[scale = 0.6]{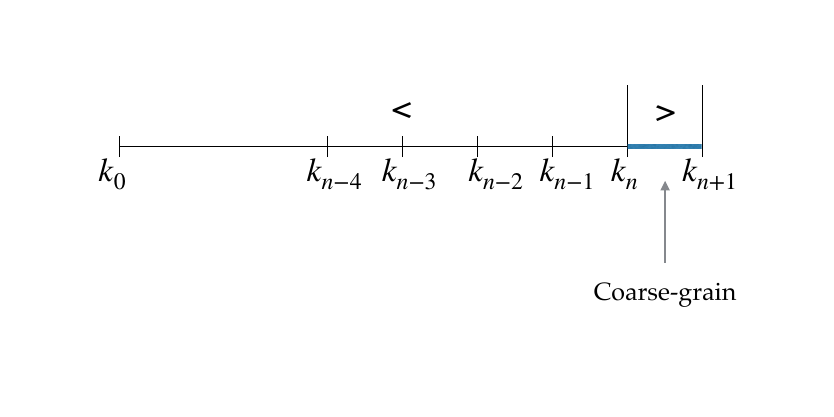}
	\end{center}
	\vspace*{0pt}
	\caption{ In wavenumber renormalization, the modes in the wavenumber band $(k_n, k_{n+1})$, denoted by $>$, are coarse-grained. The coarse-graining leads to enhancement of effective viscosity for wavenumbers $k < k_n$, denoted by $<$. }
	\label{fig:spectrum_flux}
\end{figure}

In a recursive renormalization scheme, we divide the Fourier space into wavenumber shells $(k_m, k_{m+1})$, where $k_m = k_0 b^m$ with $b>1$. We perform \textit{coarse-graining} or averaging  over a wavenumbers band, and compute its effects on the modes with lower wavenumber. Let us assume that we are at a stage with wavenumber range of $(k_0,k_{n+1})$, among which  the shell $(k_n, k_{n+1})$ is coarse-grained. See Fig.~\ref{fig:spectrum_flux} for an illustration.

We start with a dynamical equation for ${u}_1^< ({\bf k'},t)$ of Eq.~(\ref{eq:u1k_dot}). Note that ${\bf q = -k' -p}$.  The convolution in the  dynamical equation  involves  the following four sums:
\begin{align}
	[\partial_t + \nu^{(n+1)}_1 k^2 ]{u}_1^< ({\bf k'},t) = i k'  \int \frac{d{\bf p}} {(2\pi)^{d}}  \sin(\beta-\gamma) \nonumber \\ 
	[u_1^{*<}({\bf  p}, t)  u_1^{*<}({\bf q}, t) + u_1^{*<}({\bf  p}, t)  u_1^{*>}({\bf q}, t) \nonumber \\
	+  u_1^{*>}({\bf  p}, t)  u_1^{*<}({\bf q}, t)] + u_1^{*>}({\bf  p}, t)  u_1^{*>}({\bf q}, t)] 
	\label{eq:u1k_RG0}
\end{align}
because  $p $ and $q$ may be either less than $k_n$ or greater than $k_n$.  As in large-eddy simulations (LES),   $\nu^{(n+1)}_1$ in Eq.~(\ref{eq:u1k_RG0}) represents the renormalized viscosity for $u_1$ in the  wavenumber range  $(k_0, k_{n+1})$~\citep{Lesieur:book:Turbulence,Lele:JCP2003}.  Now, we ensemble-average or coarse-grain the fluctuations at scales $(k_n, k_{n+1})$. After coarse-graining, the viscosity would be $\nu^{(n)}_1$, which acts on the wavenumbers $(k_0,k_{n})$.

For the coarse-graining process,  we assume that  $u_1^{>}({\bf  k}, t) $ is  time-stationary, homogeneous, isotropic, and Gaussian with zero mean, and that $u_1^<({\bf  k}, t)$  are unaffected by coarse-graining ~\citep{Wilson:PR1974,McComb:book:Turbulence,Zhou:PR2010}.   That is,
\bea
\la u_1^{>}({\bf  k}, t) \ra & = &0, \\
\la u_1^{<}({\bf  k}, t) \ra & = & u_1^{<}({\bf  k}, t).
\eea
Therefore, assuming weak correlation between $<$ and $>$ modes, we arrive at
\bea
\la u_1^{*<}({\bf  p}, t)  u_1^{*<}({\bf q}, t) \ra & = & u_1^{*<}({\bf  p}, t)  u_1^{*<}({\bf q}, t), \\
\la u_1^{*<}({\bf  p}, t)  u_1^{*>}({\bf q}, t) \ra & = &
u_1^{*<}({\bf  p}, t)  \la u_1^{*>}({\bf q}, t) \ra = 0, \\
\la u_1^{*>}({\bf  p}, t)  u_1^{*<}({\bf q}, t) \ra & = &
\la u_1^{*>}({\bf  p}, t)  \ra u_1^{*<}({\bf q}, t) = 0 .
\eea
Substitution of the above relations in Eq.~(\ref{eq:u1k_RG0})  yields
\begin{align}
	[\partial_t + \nu^{(n+1)}_1  k^2 ]{u}^<_1({\bf k'},t) = i k'  \int \frac{d{\bf p}} {(2\pi)^{d}}  \sin(\beta-\gamma) \times  \nonumber \\ u_1^{*<}({\bf  p}, t)  u_1^{*<}({\bf q}, t)   \nonumber \\
	+ i k'  \int_\Delta \frac{d{\bf p}} {(2\pi)^{d}}  \sin(\beta-\gamma)  \la u_1^{*>}({\bf  p}, t)  u_1^{*>}({\bf q}, t) \ra ,
	\label{eq:u1k_RG1} 
\end{align}
where $\Delta$ represents the wavenumber region  $({\bf p, q}) \in (k_{n}, k_{n+1})$.  The second term of  Eq.~(\ref{eq:u1k_RG1}) enhances or renormalizes the kinematic viscosity leading to the following equation:
\begin{align}
[\partial_t + \nu^{(n)}_1  k^2 ]{u}^<_1({\bf k'},t) = i k'  \int \frac{d{\bf p}} {(2\pi)^{d}}  \sin(\beta-\gamma)
\times \nonumber \\ 
 [u_1^{*<}({\bf  p}, t)  u_1^{*<}({\bf q}, t) ],
\label{eq:u1k_RG_final} 
\end{align}
where 
\be
 \nu^{(n)}_1 k^2 =   \nu^{(n+1)}_1 k^2 - \mathrm{Second~Integral~of~Eq.~(\ref{eq:u1k_RG1})}.
\label{eq:nu_k_second_integral}
\ee

As we show below, Eq.~(\ref{eq:nu_k_second_integral}) has two solutions. The first solution corresponds to the delta-correlated $u_1$ for which the second integral of Eq.~(\ref{eq:u1k_RG1}) is trivially zero~\cite{Onsager:Nouvo1949_SH,Lee:QAM1952,	Kraichnan:JFM1973}. For this case, 
\be
 \nu^{(n)}_1 k^2 =   \nu^{(n+1)}_1 k^2 = 0
 \label{eq:nu_soln1}
 \ee
 That is, the viscosity is not renormalized, and it remains 0 at all scales. This corresponds to the absolute equilibrium solution of Euler equation that has $\nu=0$~\cite{Lee:QAM1952,	Onsager:Nouvo1949_SH,Kraichnan:JFM1973}. { Hence, Euler equation and the corresponding field-theoretic equations satisfy  time-reversal symmetry.} The second solution, which is more complex and out of equilibrium, is computed as follows.

\begin{figure}
	\begin{center}
		\includegraphics[scale = 0.55]{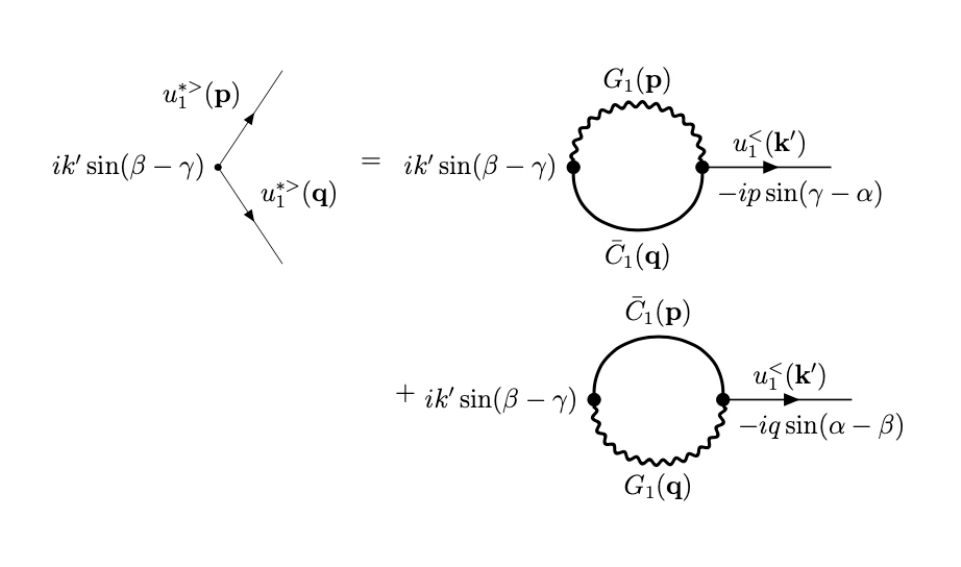}
	\end{center}
	\vspace*{0pt}
	\caption{Feynman diagrams associated with the renormalization of $\nu_1$ for the $u_1$ component.}
	\label{fig:RG_CH1}
\end{figure}

Under the quasi-gaussian approximation, the second integral of Eq.~(\ref{eq:u1k_RG1})  vanishes to the zeroth order. Hence, we expand the second term to  the first-order in perturbation that leads to   the Feynman diagrams of Fig.~\ref{fig:RG_CH1}.  We compute the integral corresponding to the first loop diagram as follows.   We expand $u_1^{*>}({\bf  p}, t)  $ using the Green's function [see Eq.~(\ref{eq:u1p_dot})]: {
\begin{align}
u_1^{*>}({\bf  p}, t) = & \int_0^t dt' G_1({\bf p},t-t') (-ip) \int \frac{d{\bf h}}{(2\pi)^d}  \sin(\gamma-\alpha) 
\times \nonumber \\ 
& u_1({\bf h},t') u_1({\bf s},t') ,
\label{eq:u1_p(t)}
\end{align}
where ${\bf p+ h + s }= 0$.}  We substitute the expression of Eq.~(\ref{eq:u1_p(t)}) in the right-hand-side of Eq.~(\ref{eq:u1k_RG1}) and simplify the expression using the following relations~\citep{McComb:book:Turbulence,Zhou:PR2010}:
\bea
\la u_1^*({\bf q}, t) u_1({\bf h}, t') \ra & = & \bar{C}_1({\bf q}, t-t')  \delta({\bf q-h}) (2\pi)^d,
\label{eq:C_q_eq_r} \\
G_1({\bf k}, t-t') & = & \theta(t-t') \exp[-\nu_1(k) k^2 (t-t') ],  \label{eq:Gk_tt'} \\
\bar{C}_1({\bf k}, t-t') & = & C_1({\bf k}) \exp[-\nu_1(k) k^2 (t-t')].  \label{eq:Ck_tt'}
\eea
In the above equations, $\bar{C}_1({\bf k}, t-t') $ is the unequal time correlation, whereas $C_1({\bf k}) $ is the equal-time correlation.  Note that $\nu_1(k)$ of Eqs.~(\ref{eq:Gk_tt'}, \ref{eq:Ck_tt'}) is the renormalized viscosity at wavenumber $k$.  As in all field theories of turbulence, we assume that the times scales for $G_1({\bf k}, t-t') $ is same as that of $\bar{C}_1({\bf k}, t-t') $.   Equation~(\ref{eq:C_q_eq_r}) yields ${\bf s = -p -h = -p -q = k'}$, using which we deduce that the  integral corresponding to the first loop diagram is
\begin{align}
I_1  = &   \int_\Delta \frac{d{\bf p}} {(2\pi)^{d}}  \int_0^t dt' G({\bf p},t-t') (k'p) \sin(\beta-\gamma)
\times \nonumber \\
& \sin(\gamma-\alpha)    \bar{C}_1({\bf q}, t-t') u^<_1({\bf k'},t'). 
\label{eq:I1_0} 
\end{align}
Now, we employ Markovian approximation~\citep{Orszag:CP1973,Leslie:book,Lesieur:book:Turbulence}. When $\nu(k) k^2 \gg 1$, the function $\exp[-\nu(k) k^2 (t-t')]$  rises sharply to unity near $t'=t$. Hence, the $dt'$ integral gets maximal contribution near $t'=t$. Therefore, $u_1({\bf k},t') \rightarrow u_1({\bf k},t)$, and
\bea
I_1 & = &   \int_\Delta \frac{d{\bf p}} {(2\pi)^{d}} \frac{kp \sin(\beta-\gamma) \sin(\gamma-\alpha) C_1({\bf q})}{\nu_1(p)p^2 + \nu_1(q) q^2} u^<_1({\bf k'},t). \nonumber \\
\label{eq:I1} 
\eea

Following similar steps, we compute the integral corresponding to the second loop diagram of Fig.~\ref{fig:RG_CH1} as
\bea
I_2 & = &    \int_\Delta \frac{d{\bf p}} {(2\pi)^{d}} \frac{kq \sin(\beta-\gamma) \sin(\alpha-\beta) C_1({\bf p})}{\nu_1(p)p^2 + \nu_1(q) q^2} u^<_1({\bf k'},t). \nonumber \\
\label{eq:I2} 
\eea
Since $I_1$ and $I_2$ are proportional to $u^<_1({\bf k'},t)$, these terms can added to $\nu^{(n+1)}_1 k^2 u^<_1({\bf k'},t)$ to yield the renormalized viscosity $\nu^{(n)}_1$. In particular, using Eqs.~(\ref{eq:u1k_RG_final}, \ref{eq:nu_k_second_integral}) we show that
\bea
\nu^{(n)}_1 k^2 & = & \nu^{(n+1)}_1 k^2 -I'_1 -I'_2 \nonumber \\
 & = & \nu^{(n+1)}_1 k^2 -  \int_\Delta \frac{d{\bf p}} {(2\pi)^{d}} \frac{k \sin(\beta-\gamma)  }{\nu_1(p)p^2 + \nu_1(q) q^2} \times \nonumber \\
 &&  [p C_1({\bf q}) \sin(\gamma-\alpha) +  q C_1({\bf p}) \sin(\alpha-\beta) ],
\nonumber \\
\label{eq:nu1(k)}
\eea
where $I'_1 , I'_2$ are $I_1,I_2$ without $u^<_1({\bf k'},t)$.

To compute $\nu^{(n)}_1$,  we choose $k=k_n$ in Eq.~(\ref{eq:nu1(k)}). In addition,   we make the following change of variables:
\be
k = k_n;~~~~~~{\bf p = p'} k_n;~~~~~~{\bf q = q'} k_n
\ee
that yields a triad $(1,p',q')$ with $1 \le p' \le b$ and $1 \le q' \le b$. We choose  $b=1.7$ for our calculation. {\citet{Zhou:NASA1997} showed that $b \in (4/3, 1.8)$ yields  a nearly constant value for the renormalized viscosity.  \citet{McComb:PRA1983}, and \citet{Zhou:PRA1988} employed $b$ in the same range. 	In our RG scheme, a  modified version of \citet{Zhou:NASA1997},  we employ $b = 1.7$ (which lies within (4/3,1.8)) so that the renormalized parameter  and Kolmogorov’s constant are close to the experimental values.}

 For the integral we employ $p'$ and $z = \cos\gamma$, where $\gamma$ is the angle between ${\bf k}$ and ${\bf p}$, as the independent variables that yields
\be
\int d{\bf p} = S_{d-1} \int_\Delta p'^{d-1} dp' \int_{\Delta'} dz (1-z^2)^{\frac{d-3}{2}}
\ee
where $\Delta, \Delta'$ are the domain of integrations: $p'=[1,b]$ and $z = [(p'^2 + 1 -b^2)/(2p'), p'/2]$, in which the latter limits are obtained by setting $q'=(1,b)$.  In this paper, we focus on $k^{-5/3}$ spectral regime, for which $C_1({\bf k}) $ is given by Eq.~(\ref{eq:Ek_Ck}), and
\bea
E(k) & = & K_\mathrm{Ko} \epsilon_u^{2/3} k^{-5/3},  \label{eq:Ek_2d} \\
\nu^{(n)}_1 & = & \nu_{1*} \sqrt{K_\mathrm{Ko}} \epsilon_u^{1/3} k_n^{-4/3},  \label{eq:nuk_2d}
\eea
where $\epsilon_u$ is the energy flux, $K_\mathrm{Ko}$ is the Kolmogorov constant, and $\nu_{1*}$ is the renormalization constant for $u_1$.  

We substitute Eqs.~(\ref{eq:Ek_2d}, \ref{eq:nuk_2d}) in   Eq.~(\ref{eq:nu1(k)}), and simplify the expressions using trignometric identities for the  triad $(1,p',q')$ (see Fig.~\ref{fig:CH_triad}). 
At $k=k_n$, these operations yield
\begin{align}
\nu_{1*}  (1- b^{-4/3})  + &  \frac{2 S_{d-1}}{(d-1) S_d} \frac{1}{\nu_{1*}} \int_1^b p'^{d-1} dp' \nonumber \\ \int^{p'/2}_{(p'^2+1-b^2)/(2p')}  dz  &(1-z^2)^{\frac{d-3}{2}} (F_1+F_2) = f(\nu_{1*} )=0,
\label{eq:nu1_integral}
\end{align}
where  
\bea
F_1(p',z) & = & \frac{(1-z^2)  (p'-2z)(2p'z-1) p' q'^{-8/3-d}}{p'^{2/3} +q'^{2/3}}, \\
F_2(p',z)  & = & \frac{(1-z^2)  (1-p'^2)(2p'z-1) p'^{-2/3-d} q'^{-2}}{p'^{2/3} +q'^{2/3}}
\eea
are functions of the independent variables $p'$ and $z$.   Equation~(\ref{eq:nu1_integral}) differs from those employed by \citet{McComb:PRA1983} and \citet{Zhou:PRA1989} who computed the correction to $\nu_1(k)$ [Eq.~(\ref{eq:nu1(k)})] for all $k$'s that leads to a $k$-dependent $\nu_{1*}$.  In our paper, we interpret $\nu_1^{(n)}$ as the renormalized viscosity for wavenumbers $(k_0, k_n)$ that leads to a constant $\nu_{1*}$.  Our scheme, which is motivated by LES~\citep{Lesieur:book:Turbulence,Lele:JCP2003}, 
simplifies the computation of $\nu_{1*}$ significantly. 

\begin{figure}
	\begin{center}
		\includegraphics[scale = 0.8]{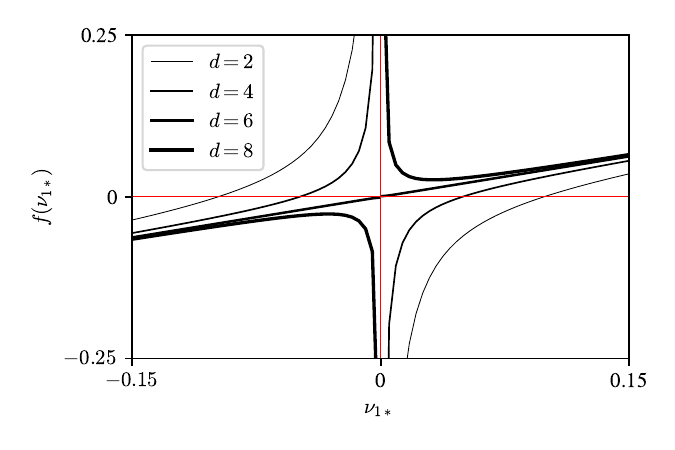}
	\end{center}
	\vspace*{0pt}
	\caption{Plot of $f(\nu_{1*})$ vs.~$\nu_{1*}$ [Eq.~(\ref{eq:nu1_integral})]. The solution $\nu_{1*}$ is the  root of  $f(\nu_{1*})=0$. We have finite $\nu_{1*}$ (both positive and negative) for $d<6$, but it has no solution for $d \ge 6$. For $d>6$, the allowed solution for the RG equation is $\nu_{1*} = 0$.}
	\label{fig:f_nu1}
\end{figure}

The solution $\nu_{1*}$ is the root of $f(\nu_{1*}) = 0$ [see  Eq.~(\ref{eq:nu1_integral})], which is illustrated in  Fig.~\ref{fig:f_nu1} for $d=2,4,6,8$.  For $d<6$, we have positive and negative roots, out of which  only the positive root is sensible because it leads to diminishing temporal correlation with the increase of $t-t'$ [see Eq.~(\ref{eq:Ck_tt'})] and negative energy flux for $d=2$ in the $k^{-5/3}$ regime. Hence, we work with positive $\nu_{1*}$ for $d<6$.  In Table~\ref{tab:nu_Ko} we list $\nu_{1*}$ for various $d$'s. Note that $\nu_{1*}$ decreases gradually to zero as   $d \rightarrow 6$. 

For $d \ge 6$, Eq.~(\ref{eq:nu1_integral}) has no root. Therefore,  $\nu_{1*} = 0$,  the equilibrium solution of Euler equation, is the only solution for the RG equation. This is similar to Wilson's $\phi^4$ theory~\cite{Wilson:PR1974},
 where the system transitions from nontrivial fixed point to guassian fixed point at $d=4$.  These observations indicate that $d=6$ is  the \textit{upper critical dimension}.  

\begin{figure}
	\begin{center}
		\includegraphics[scale = 0.9]{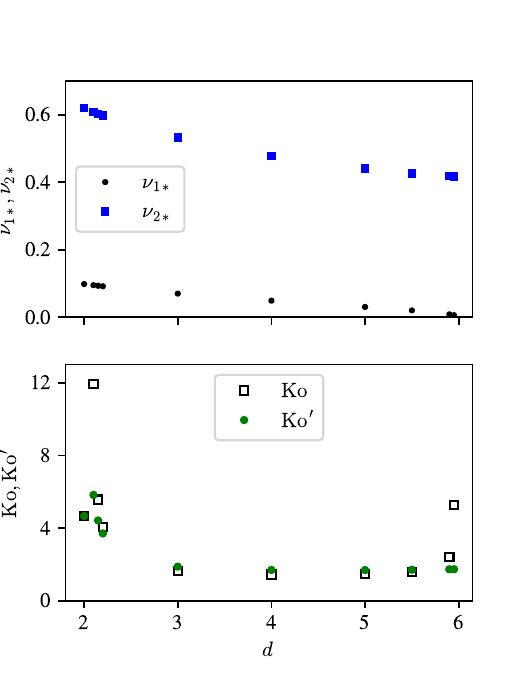}
	\end{center}
	\vspace*{0pt}
	\caption{Values of constants for various $d$'s: (a) $\nu_{1*}$ and $\nu_{2*}$; (b) $\mathrm{Ko}$ and $\mathrm{Ko}'$.}
	\label{fig:nu_Ko}
\end{figure}

To determine $\nu_{1*}$, we compute the integral of Eq.~(\ref{eq:nu1_integral}) numerically.  For    an accurate  integration, we perform the $dz$ integral  using Gaussian quadrature and the  $dp'$ integral  using a Romberg scheme.  In addition, we employ \textit{mid-point method} for computing the  roots of Eq.~(\ref{eq:nu1_integral}). Refer to Appendix~\ref{sec:integration} for  details on the integration schemes used in this paper.  We employ Python's {\tt scipy.integrate.romberg} function   whose tolerance limit is $ 1.48 \times 10^{-8} $.

\begin{table}[!h]
	\begin{center}
		\caption{Table showing the renormalization constants, $\nu_{1*}$ and $\nu_{2*}$, and Kolmogorov's constants, $\mathrm{Ko}$ and $\mathrm{Ko}'$.  For $d=6$,  the parameters correspond to the equilibrium solution. }
		\label{demo-table}
		\begin{tabular}{c c c c c} 
			\hline
			$d$ & $\nu_{1*}$ & $\nu_{2*}$ & $\mathrm{Ko}$ & $\mathrm{Ko}'$ \\ [0.5ex] 
			\hline\hline
			2 &  0.098 & 0.619 & 4.66 &  4.66 \\
			\hline
			2.1 & 0.095 & 0.608 &  11.93  & 5.82 \\
			\hline
			2.15 & 0.093 & 0.603 & 5.56 & 4.42  \\
			\hline
			2.2 & 0.092 & 0.598 & 4.05 & 3.70  \\
			\hline
			3 & 0.070 & 0.533 & 1.63 & 1.88 \\
			\hline
			4 & 0.049 & 0.479 & 1.45 & 1.69 \\
			\hline
			5 & 0.030 & 0.441 &  1.48 & 1.69 \\
			\hline
			5.95 & 0.006 & 0.417 & 5.27 & 1.73 \\ 	
			\hline
			6* & 0 & 0 & - & -  \\ [1ex] 
			\hline
		\end{tabular}
			\label{tab:nu_Ko}
	\end{center}
\end{table}

For $d=2$, $\nu_{1*} $ is the only renormalized parameter. However, higher dimensions have both  $\nu_{1*}$ and  $\nu_{2*}$. Refer to \citet{Verma:arxiv2023} for a detailed comparison of our $\nu_{1*}$ with those reported earlier.  In the next subsection, we will compute the renormalized viscosities for the $u_2$, ..., $u_{d-1}$ components.

\subsection{Renormalization of  $u_j$ ($j>2$) Components}
\label{sec:RG_u2}

For isotropic turbulence, the renormalized viscosities for the components $u_2$, ..., $u_{d-1}$  are the same. We denote this quantity as   $\nu_2^{(n)}$ and compute it following the same steps as in Section~\ref{sec:RG_u1}, but with Eq.~(\ref{eq:u2k_dot}). One of the intermediate steps in the derivation of  $\nu_2^{(n)}$ is 
\begin{align}
& (\partial_t + \nu_2^{(n+1)} k^2){u}^<_j({\bf k'},t)  = 
\nonumber \\
& i k' \int \frac{d{\bf p}} {(2\pi)^{d}}  \{ \sin \gamma u_1^{<*}({\bf p},t)  u_j^{<*}({\bf q},t) -\sin\beta u_1^{<*}({\bf q},t)  u_j^{<*}({\bf p},t)\} \nonumber \\
& + i k' \int \frac{d{\bf p}} {(2\pi)^{d}}   \{ \sin \gamma u_1^{>*}({\bf p},t)  u_j^{>*}({\bf q},t) \nonumber \\ 
& ~~~~~~~~~-\sin\beta u_1^{>*}({\bf q},t)  u_j^{>*}({\bf p},t)\} ,
 \label{eq:u2k_RG} 
\end{align}
where  $j\ge 2$. In the above equation, the terms of the form $\int d{\bf p} \la u_1^{>*}({\bf q},t)  u_j^{>*}({\bf p},t) \ra$ contribute to viscosity renormalization.  In this subsection we show that $\nu_2^{(n+1)} \ne \nu_1^{(n+1)}$, which is expected because  $u_1$ and $u_j$ with $j \ge 2$  evolve differently [Eqs.~(\ref{eq:u1k_dot}, \ref{eq:u2k_dot})].

\begin{figure}
	\begin{center}
		\includegraphics[scale = 0.5]{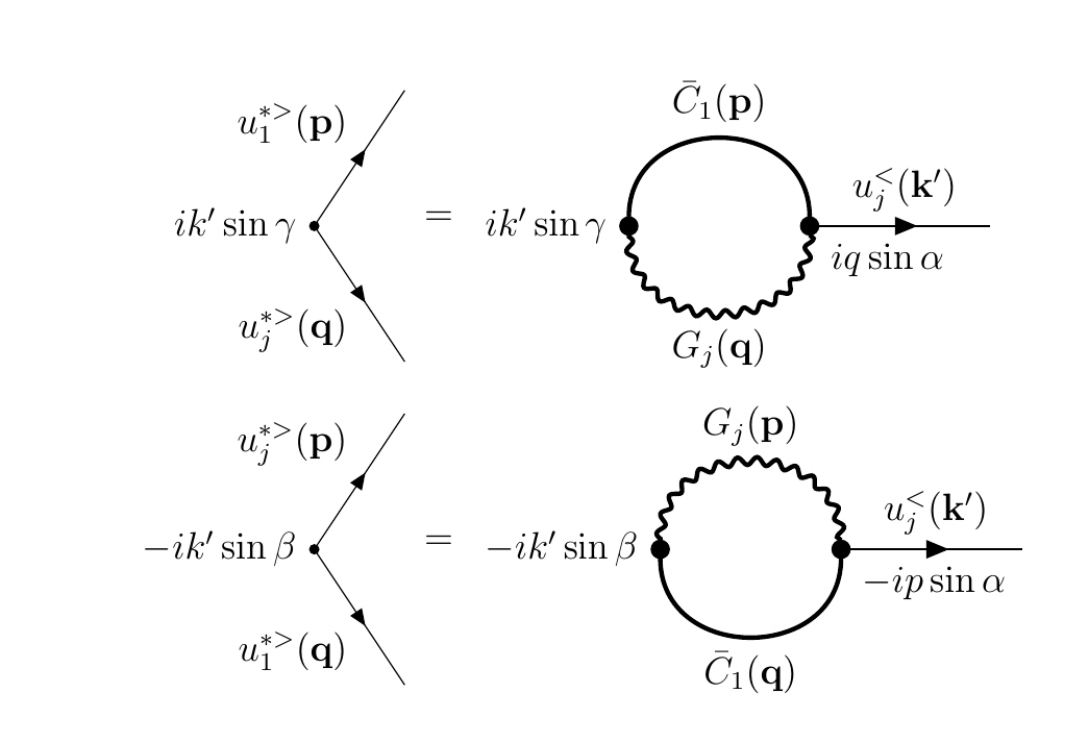}
	\end{center}
	\vspace*{0pt}
	\caption{Feynman diagrams associated with the renormalization of $\nu_2$ for the $u_j$ component. }
	\label{fig:RG_CH2}
\end{figure}

As in Section~\ref{sec:RG_u1}, we employ the isotropic correlation function of Eq.~(\ref{eq:Ek_Ck}) and 
\bea
\nu^{(n)}_2 & = & \nu_{2*} \sqrt{K_\mathrm{Ko}} \epsilon_u^{1/3} k_n^{-4/3}, \label{eq:nu2_defn}
\eea
where $\epsilon_u$ is the energy flux, and $\nu_{2*} $ is the renormalization constant for $\nu_j$ with $j\ge 2$. 
The second integral of Eq.~(\ref{eq:u2k_RG}) contributes to the viscosity renormalization, whose associated Feynman diagrams are shown in Fig.~\ref{fig:RG_CH2}, and the corresponding integral is
\bea
I_3 & = &   - u^<_2({\bf k'},t) \int_\Delta \frac{d{\bf p}} {(2\pi)^{d}} 
\left[ \frac{kq C_1({\bf p}) \sin \gamma \sin \alpha 
}{\nu_1(p)p^2 + \nu_2(q) q^2} \right.  \nonumber \\
&& \left.  ~~~~~~~+ \frac{ kp C_1({\bf q}) \sin \beta \sin \alpha 
}{\nu_2(p)p^2 + \nu_1(q) q^2}  \right]
\label{eq:I3} 
\eea
that contributes to  the viscosity renormalization    as follows:
\bea
\nu_{2*}  (1-b^{-4/3}) & = & - \frac{2 S_{d-1}}{(d-1) S_d} \int_1^b p'^{d-1} dp' \nonumber \\
&&  \int^{p'/2}_{(p'^2+1-b^2)/(2p')}    dz (1-z^2)^{\frac{d-3}{2}}   F_3(p',z) ,\nonumber \\
\label{eq:nu2_integral}
\eea
with
\bea
F_3(p',z) & = &  \frac{(1-z^2)  p'^{-2/3-d}}{\nu_{1*} p'^{2/3} + \nu_{2*} q'^{2/3}}
	+ \frac{(1-z^2) p'^2 q'^{-8/3-d}}{\nu_{2*} p'^{2/3} + \nu_{1*} q'^{2/3}}. \nonumber \\
\eea

We solve for $\nu_{2*}$ by iterating Eq.~(\ref{eq:nu2_integral}) starting with a guess value of $\nu_{2*}$. The iterative process converges to $\nu_{2*}$  listed in Table \ref{tab:nu_Ko} and illustrated in Fig.~\ref{fig:nu_Ko}. Since nonzero $\nu_{1*}$  solution exists only for $d<6$, Eq.~(\ref{eq:nu2_integral}) implies that $\nu_{2*}$ too is valid for $d<6$. For $d \ge 6$,  $\nu_{1} = \nu_{2}=0$ that corresponds to the equilibrium solution of Euler equation.  Note $\nu_{1*} \ll \nu_{2*}$, as illustrated in Fig.~\ref{fig:nu_Ko}.

For $d \ge 3$, hydrodynamic turbulence exhibits multitudes of triads, each of which have different $\hat{n}$.  Hence, for a given ${\bf k}$, the Craya-Herring vectors of Fig.~\ref{fig:CH_triad} transform to each other (depending on the triads).  Note, however, that $\nu_{1}({\bf k}) \ll \nu_{2}({\bf k})$, hence we may estimate that $\nu(k) \approx \nu_2(k)$, which would be useful  for LES.  {In spite of the above complications,   independent  evaluations of  $\nu_{1}({\bf k})$ and $ \nu_{2}({\bf k})$ yield valuable insights, chiefly that $\nu_{1*} \rightarrow 0$ as $d \rightarrow 6$, leading to the upper critical dimension of hydrodynamic turbulence as 6. The earlier works, e.g., \cite{Fournier:PRA1978},  could not reach this result because they did not resolve the renormalized viscosities for the different components of the velocity field. }

Using Eqs.~(\ref{eq:nuk_2d}, \ref{eq:nu2_defn}), we derive that for both $\nu_1^{(n)}$ and $\nu_2^{(n)}$, 
\be
\frac{\nu_{1,2}^{(n)}}{\nu_{1,2}^{(n)}}
= \left( \frac{k_{n+1}}{k_n} \right)^{-4/3} =  b^{-4/3} .
\ee
In 	quantum field theory,  we express the running coupling constant in terms of $b = \exp(l)$~\citep{Peskin:book:QFT}.  { Using $b^{-4/3} \approx 1- 4l/3$ (for small $l$), we derive the beta function for $\nu$ using
\be
\frac{d \nu}{dl} \approx - \frac{4}{3} \nu,
\ee
or
\be
\beta(\nu) = \frac{d \log \nu}{d \log k} \approx - \frac{4}{3}.
\ee
Note that the beta function for the coupling constant is
\be
\beta(\lambda) = \frac{d \log \lambda}{d \log k} =0
\ee
due to Galilean invariance.  These relations would be useful in relating field theory of turbulence and quantum field theory~\cite{Peskin:book:QFT}.}


In the next section, we compute the energy flux using field theory.

\section{Energy Transfers and Fluxes in $d$ dimensions}
\label{sec:ET}

In this section, we compute the energy transfer rates and energy flux in the inertial range of hydrodynamic turbulence.  

\subsection{Energy Transfers and Flux  for  $u_1$ Component }
\label{sec:M2M_CH1}

\begin{figure}
	\begin{center}
		\includegraphics[scale = 0.30]{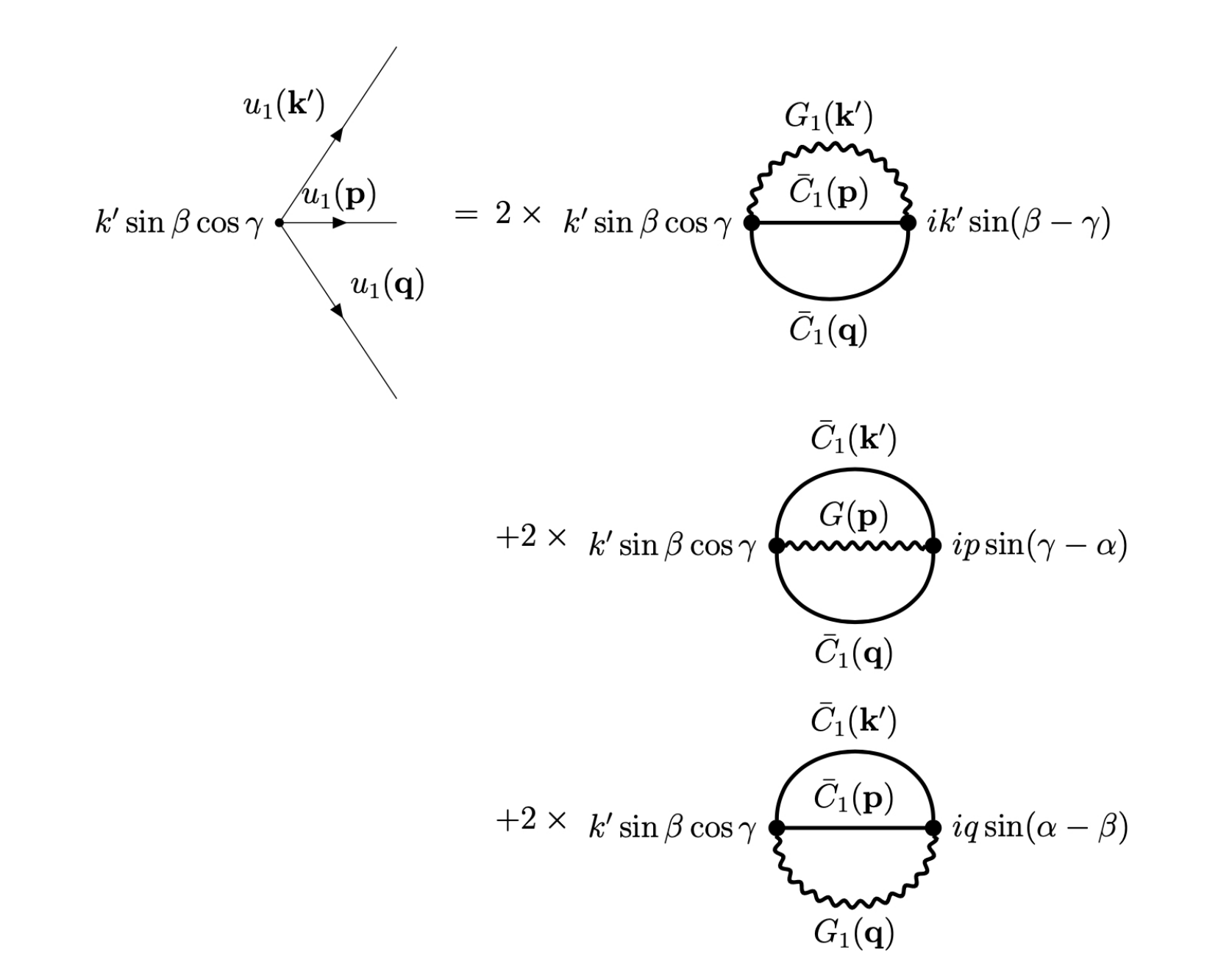}
	\end{center}
	\vspace*{0pt}
	\caption{Feynman diagrams associated with the energy transfers between the $u_1$ components.  Equation~(\ref{eq:four_corr_QN}) illustrates two ways to get the second-order correlation functions, which leads to the factor 2 in all the digrams. }
	\label{fig:ET_CH1}
\end{figure}

In this subsection, we will compute the mode-to-mode energy transfers among   the $u_1$ components within a triad.  We start with Eq.~(\ref{eq:Su1u1}) and present the ensemble-averaged \textit{mode-to-mode energy transfer} from $u_1({\bf p},t)$ to $u_1({\bf k'},t)$  with the mediation of $u_1({\bf q},t)$, which is
\bea
\la S^{u_1 u_1}({\bf k'|p|q}) \ra & = &k' \sin\beta \cos \gamma  \times \nonumber \\
&&  \Im \{ \la u_1({\bf q},t) u_1({\bf p},t) u_1({\bf k'},t) \ra \}   
\label{eq:Skpq_u1_avg}
\eea
with ${\bf k' + p+q } =0$.   Following earlier literature~\cite{Kraichnan:JFM1959,	Orszag:CP1973}, we assume that the variables $u_1({\bf p},t)$, $u_1({\bf k'},t)$, and $u_1({\bf q},t)$ are quasi-normal.  Under this assumption, the  triple correlation of Eq.~(\ref{eq:Skpq_u1_avg}) vanishes to the zeroth order. However,  the first-order expansion of the triple correlation of Eq.~(\ref{eq:Skpq_u1_avg}) leads to a fourth-order correlation, that is expanded as a sum of  products of two second-order correlations.  The corresponding Feynman diagrams are given in  Fig.~\ref{fig:ET_CH1}.

\begin{figure*}
	\begin{center}
		\includegraphics[scale = 0.9]{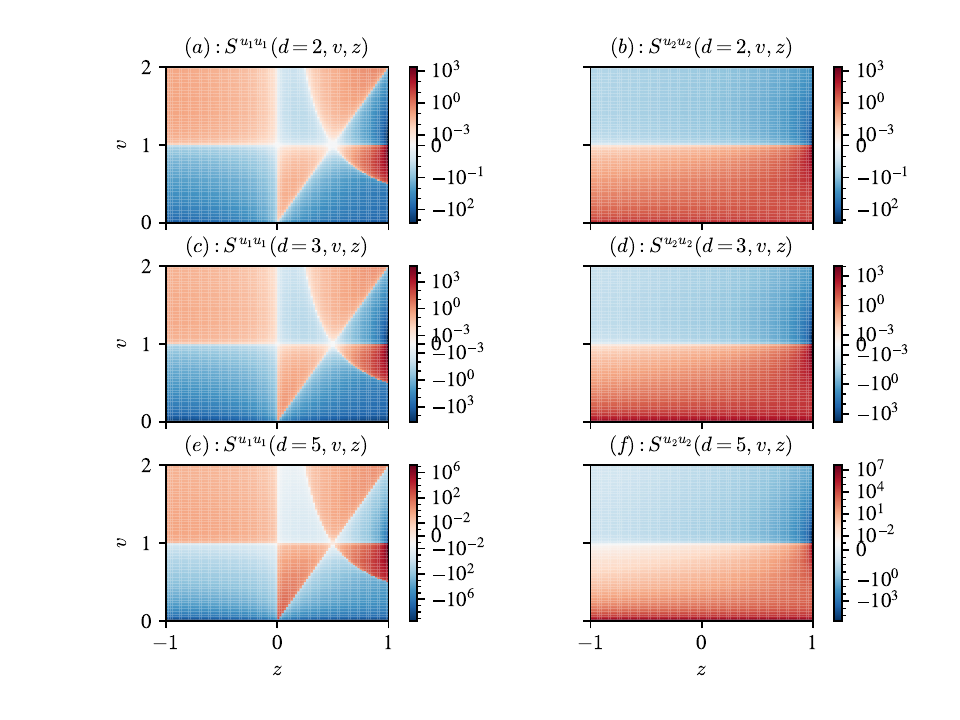}
	\end{center}
	\vspace*{0pt}
	\caption{The density plots of  $\la S^{u_1 u_1}(v,z) \ra $ and $\la S^{u_2 u_2}(v,z) \ra $  for $d=2$ (top row), $d=3$ (middle row), and $d=5$ (bottom row).}
	\label{fig:Skpq}
\end{figure*}

Let us evaluate the integral corresponding to the first Feynman diagram of Fig.~\ref{fig:ET_CH1}. Here, $u_1({\bf k'},t) $ is expanded using the Green's function as [see Eq.~(\ref{eq:u1k_dot})]
\bea
u_1({\bf k'},t) & = & i \int_0^t dt' G_1({\bf k'},t-t') 
 k'  \int  \frac{d{\bf h}}{(2\pi)^d}   \sin(\beta-\gamma) \times \\
 && [u_1^*({\bf  h}, t')  u_1^{*}({\bf s}, t') ]
\eea
with ${\bf k' + h+s } =0$. Substitution of the above in Eq. (\ref{eq:Skpq_u1_avg}) leads to a fourth-order correlation, which is expanded as a sum of products of two second-order correlations:
\begin{align}
& \la u_1({\bf q},t) u_1({\bf p},t) u_1({\bf h},t') u_1({\bf s},t') \ra \nonumber  \\
& = \la u_1({\bf q},t) u_1({\bf p},t) \ra \la  u_1({\bf h},t') u_1({\bf s},t') \ra  \nonumber  \\
& +  \la u_1({\bf q},t)  u_1({\bf h},t') \ra \la u_1({\bf p},t) u_1({\bf s},t') \ra \nonumber  \\
& + \la u_1({\bf q},t) u_1({\bf s},t')  \ra \la u_1({\bf p},t) u_1({\bf h},t') \ra.
  \label{eq:four_corr_QN}
\end{align}
Note that $\la u_1({\bf q},t) u_1({\bf p},t) \ra  = \la  u_1({\bf h},t') u_1({\bf s},t') \ra = 0$  because ${\bf p +  q = k} \ne 0$ and ${\bf r +  s = k} \ne 0$. 
Using the above correlations, we deduce that
\bea
\la u_1({\bf q},t) u_1({\bf p},t) u_1({\bf k'},t) \ra_a  
& = &  \int_0^t dt' G_1({\bf k'},t-t') 
i k'  \times \nonumber \\
&&   \sin(\beta-\gamma)  2 \bar{C}_1({\bf  p},t- t')  \bar{C}_1({\bf q}, t-t')  . \nonumber \\
\eea
Using the properties of temporal relations of Eqs.~(\ref{eq:Gk_tt'}, \ref{eq:Ck_tt'}), we deduce that
\bea
\la u_1({\bf q},t) u_1({\bf p},t) u_1({\bf k'},t) \ra  
& = &   \frac{i 2  k' \sin(\beta-\gamma) C_1({\bf p}) C_1({\bf q})}{\nu_1(k) k^2 + \nu_1(p) p^2 + \nu_1(q) q^2}. \nonumber \\
\eea
This term plus other two terms of Fig.~\ref{fig:ET_CH1}  yields 
\bea
\la S^{u_1 u_1}({\bf k'|p|q}) \ra & = &     \frac{\mathrm{numr}_1 }{\nu_1(k) k^2 + \nu_1(p) p^2 + \nu_1(q) q^2},
\label{eq:Skpq_u1_expanded}
\eea
where
\bea
\mathrm{numr}_1 & = & 2 k' \sin\beta \cos \gamma  [k' \sin(\beta-\gamma) C_1({\bf p})C_1({\bf q})
\nonumber \\
&& +p \sin(\gamma-\alpha) C_1({\bf k'} ) C_1({\bf q}) \nonumber \\
&& 
+q \sin(\alpha-\beta) C_1({\bf k'} ) C_1({\bf p}) ] .
\eea
The physics in the inertial range is scale invariant, hence we  employ  the following transformations~\cite{Kraichnan:JFM1959,Leslie:book}:
\be
k = \frac{R}{u};~~~p = \frac{Rv}{u};~~~q = \frac{Rw}{u};
\label{eq:kpq_transform}
\ee
 that  leads to
\bea
\la S^{u_1 u_1}({\bf k'|p|q}) \ra & = &      
(2\pi)^{2d}  \epsilon_u k^{-2d}
K_\mathrm{Ko}^{3/2} \frac{4}{S_d^2(d-1)^2} \la S^{u_1 u_1}(v,z) \ra, \nonumber \\
\label{eq:Svw_u1}
\eea
where
\bea
\la S^{u_1 u_1}(v,z) \ra & = &     \frac{\mathrm{numr}_2 }{ \nu_{1*} (1+v^{2/3} + w^{2/3})} 
\label{eq:Svw_u1b}
\eea
with
\bea
\mathrm{numr}_2 & = &2 (2vz-1)(1-z^2) zv  w^{-2}  (vw)^{-2/3-d}  \nonumber \\
&& + 2 (v-2z)(1-z^2) zv^2  w^{-8/3-d}  \nonumber \\
&& + 2 (1-v^2)(1-z^2)  z  w^{-2} v^{-1/3-d}
\label{eq:numr_Skpq_2D} 
\eea
and $w^2 = 1+ v^2 - 2 v z$.   Note that 
$\la S^{u_1 u_1}({\bf k'|p|q}) \ra$ has a  dimension of $k^{-2d}$, whereas  $\la S^{u_1 u_1}(v,z) \ra$ is dimensionless.  In Fig.~\ref{fig:Skpq}(a,c,e) we illustrate the density plots of $\la S^{u_1 u_1}(v,z) \ra$ for $d =2,3,5$ respectively.  

As is evident in Fig.~\ref{fig:Skpq}(a,c,e), the function $\la S^{u_1 u_1}(v,z) \ra$ exhibits the following interesting properties:
\begin{enumerate}
	\item  $\la S^{u_1 u_1}(v,z) \ra \rightarrow \infty $ as   $z \rightarrow 1$ and $v \rightarrow 1$, hence $\la S^{u_1 u_1}(v,z) \ra$  is singular near this region. Note that $\la S^{u_1 u_1}(v,z) \ra$ in the figure is bounded  due to the finite resolution.   For $z\approx 1$, $\la S^{u_1 u_1}(v,z) \ra$ is positive when $v < 1$ and negative otherwise. This feature illustrates forward energy transfers in hydrodynamic turbulence~\cite{Domaradzki:PF1990,Verma:Pramana2005S2S,Verma:arxiv2023}.
	
	\item $\la S^{u_1 u_1}(v,z) \ra$ takes large negative values for all $z$'s when $v \rightarrow 0$. These are nonlocal reverse energy transfers from $k'$ to $p$ when $p \ll k'$. These transfers are responsible for the inverse energy cascade in 2D hydrodynamic turbulence when $E(k) \sim k^{-5/3}$.
	
	\item The singularity of $\la S^{u_1 u_1}(v,z) \ra$   become more severe with the increase of $d$. For example, asymptotically  $\la S^{u_1 u_1}(v,z) \ra \rightarrow (1-v)^{-8/3-d}$ as $v \rightarrow 1$ and $z \rightarrow 1$~\cite{Verma:Pramana2005S2S,Verma:arxiv2023}.
\end{enumerate}

The above properties are in agreement with the earlier results~\cite{Verma:Pramana2005S2S,Verma:arxiv2023}.

After a brief discussion on   $\la S^{u_1 u_1}(v,z) \ra$ we compute the energy flux in 2D  arising from the  cumulative energy transfers:
\be
\la \Pi_{u_1}(R) \ra = \int_{R}^\infty \frac{d{\bf k'}}{(2\pi)^2} \int_0^{R} \frac{d{\bf p}}{(2\pi)^2} \la S^{u_1 u_1}({\bf k'|p|q}) \ra  .
\label{eq:Pi_2d}
\ee
Substitution of Eq.~(\ref{eq:Svw_u1}) transforms the energy flux equation   to~\cite{Kraichnan:JFM1959,Verma:arxiv2023}
\bea
\frac{ \la \Pi_{u_1}(R)\ra}{\epsilon_u } &  = &  A  \int_0^1 dv   [\log(1/v)] v^{d-1} \int_{-1}^1 dz (1-z^2)^{\frac{d-3}{2}} \nonumber \\
&& \la S^{u_1 u_1}(v,z) \ra, 
\label{eq:Pi_2D_integral}
\eea
where
\be
A =  K_\mathrm{Ko}^{3/2}
\frac{4}{(d-1)^2} \frac{S_{d-1}}{S_d} .
\label{eq:A}
\ee
We compute the double integral of Eq.~(\ref{eq:Pi_2D_integral}) numerically. We employ  Gauss-Jacobi quadrature for the $dz$ integral, and  Romberg iterative scheme for the $dv$ integral. Refer to Section~\ref{sec:integration} for a brief discussion on the integration procedure.  

{ Two-dimensional hydrodynamics has only  $u_1$ component. Hence, $\Pi_{u_1}(R)$ is the energy flux for 2D turbulence. In the $k^{-5/3}$ regime of 2D turbulence, we observe that $\Pi_{u_1}(R) < 0$ indicating an inverse cascade of energy, consistent with the predictions of \citet{Kraichnan:JFM1971_2D3D}.  Using $\Pi_{u_1}(R) = -\epsilon_u$ and $\nu_{1*} =  0.098$, we deduce that $K_\mathrm{Ko} = 1.19$.  
This $K_\mathrm{Ko} $  is lower than that reported in experiments and numerical situations, which is approximately 6.  This inconsistency is possibly due to the inability of the recursive RG schemes to capture the nonlocal interactions~\cite{Verma:arxiv2023}. Note that 2D hydrodynamic turbulence involves local forward energy transfer and nonlocal inverse energy transfer, which is difficult to incorporate in RG procedure.   \citet{Verma:arxiv2023} employed a temporary fix for this problem by increasing the lower cutoff of the flux integral to 0.22. We find that $\int_{0.22}^1 dv ... $  yields $K_\mathrm{Ko} = 4.46$, which is close to the  earlier numerical and experimental results.  Hopefully,  in future we will understand the reason for the cutoff better.}

\subsection{Energy Transfers and Fluxes  for the $u_j$  Components with $j\ge 2$}
\label{sec:ETu2}
\begin{figure}
	\begin{center}
		\includegraphics[scale = 0.5]{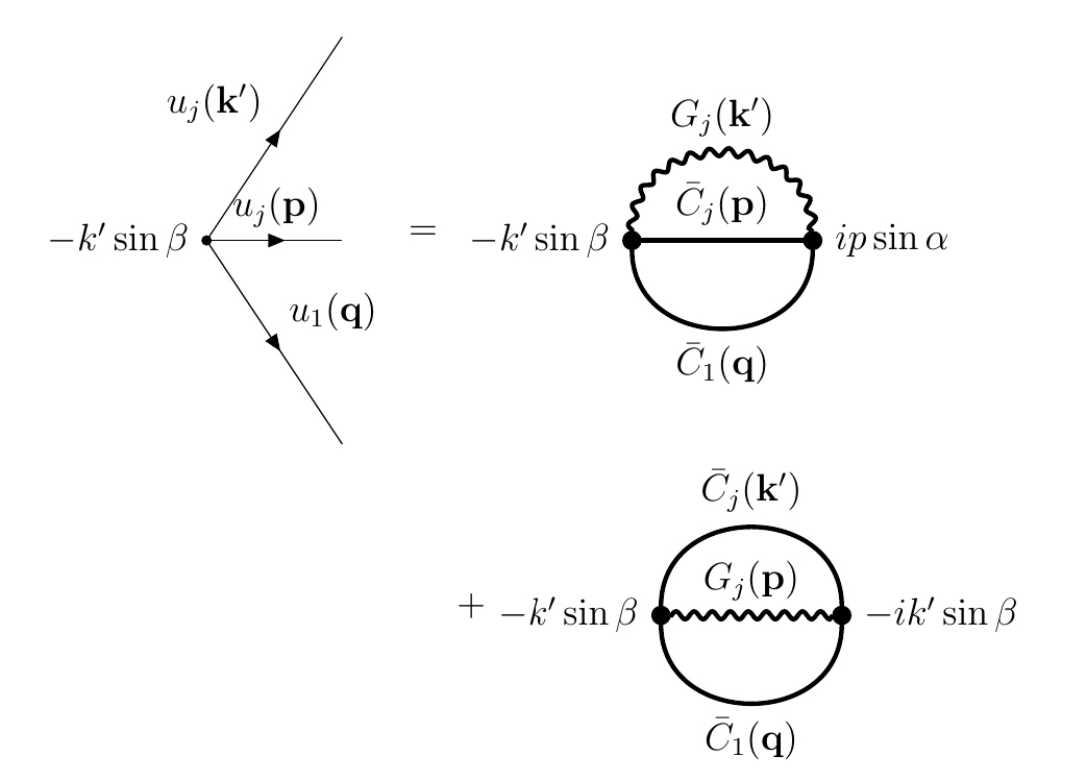}
	\end{center}
	\vspace*{0pt}
	\caption{Feynman diagrams associated with the energy transfers between the $u_j$ component ($j \ge 2$). }
	\label{fig:ET_CH2}
\end{figure}

As shown in Eq.~(\ref{eq:Su2u2}), the mode-to-mode energy transfer from $u_j({\bf p})$ to $u_j({\bf k'})$ [$j \ge 2$] with the mediation of $u_1({\bf q})$  is 
\bea
\la S^{u_j u_j}({\bf k'|p|q}) \ra  & = & - k' \sin\beta \Im \{ \la u_1({\bf q},t) u_j({\bf p},t) u_j({\bf k'},t) \ra \}   \nonumber \\.\label{eq:Su2u2_avg}
\eea
 Note that $\la S^{u_j u_j}({\bf k'|p|q}) \ra$ are the same for all $j$'s from $j=2$ to $d-1$ due to isotropy.
 To compute this quantity we employ the  scheme described in Sec.~\ref{sec:M2M_CH1}.  For simplicity, we restrict ourselves to  flows for which $\la u_1 u_j \ra =0$ when $j > 1$. Consequently,   the expansion of $u_j$ components in terms of the Green's function yields nonzero values, whereas the terms arising  from the expansion of $u_1$ component vanishes identically.    The  Feynman diagrams associated with   $\la S^{u_j u_j}({\bf k'|p|q}) \ra$ (for $j\ge 2$) are illustrated in Fig.~\ref{fig:ET_CH2}.

Following the same steps as in Section~\ref{sec:M2M_CH1},
the field-theoretic estimate  for  $\la S^{u_j u_j}({\bf k'|p|q}) \ra $ is
\bea
\la S^{u_j u_j}({\bf k'|p|q}) \ra  & = & (k \sin\beta)^2   \frac{C_1({\bf q}) [C_j({\bf p}) -C_j({\bf k'})  ] }{\nu_2(k) k^2 + \nu_2(p) p^2 + \nu_1(q) q^2}.  \nonumber \\
\eea
We assume that turbulence is isotropic, hence $ C_j({\bf k'}) = C({\bf k'})$.  In addition, we transform $ \la S^{u_j u_j}({\bf k'|p|q}) \ra$ to $\la S^{u_j u_j}(v,z) \ra$ as follows:
\bea
\langle S^{u_j u_j}({\bf k'|p|q}) \rangle & = &
(2\pi)^{2d}  \epsilon_u k^{-2d}
{K_\mathrm{Ko}^{3/2}} \frac{4}{S_d^2(d-1)^2} 
 \nonumber \\
 && \langle S^{u_j u_j}(v,z) \rangle
\eea
with
\bea
\la S^{u_j u_j}(v,z) \ra & = &     \frac{v^2 w^{-8/3-d} (v^{-2/3-d}-1) (1-z^2)}{ \nu_{2*}(1+v^{2/3}) +\nu_{1*}  w^{2/3}}.
\label{eq:SCH2(v,w)}
\eea
 
In Fig.~\ref{fig:Skpq}(b,d,f), we illustrate $\la S^{u_2 u_2}(v,z) \ra$ for $d=2,3,5$. As shown in the figure, $\la S^{u_2 u_2}(v,z) \ra$ exhibits the following interesting properties:
\begin{enumerate}
	\item  $\la S^{u_j u_j}(v,z) \ra$  is singular when $z \rightarrow 1$ and $v \rightarrow 1$.    $\la S^{u_j u_j}(v,z) \ra$  is positive when $v < 1$ and negative otherwise, implying  forward energy transfers when $p \rightarrow k'$~\cite{Domaradzki:PF1990,Verma:Pramana2005S2S,Verma:arxiv2023}.
	
	\item $\la S^{u_j u_j}(v,z) \ra \gg 1$ for all $z$'s when $v \rightarrow 0$. These transfers represent forward nonlocal energy transfers from $p$ to $k'$.
	
	\item The severity of the singularity of $\la S^{u_j u_j}(v,z) \ra$  increases with  $d$, with   $\la S^{u_2 u_2}(v,z) \ra \rightarrow (1-v)^{-8/3-d}$ as $v \rightarrow 1$ and $z \rightarrow 1$~\cite{Verma:Pramana2005S2S,Verma:arxiv2023}.
\end{enumerate}

Now we compute the energy flux that receives contributions from  the $u_j$ components  as given below:
\be
\la \Pi_{u_j}(R) \ra = \int_{R}^\infty \frac{d{\bf k'}}{(2\pi)^3} \int_0^{R} \frac{d{\bf p}}{(2\pi)^3}    \la S^{u_j u_j}({\bf k'|p|q}) \ra,
\ee
hence 
\bea
\frac{ \la \Pi_{u_j}(R)\ra}{\epsilon_u }  & = & A  \int_0^1 dv   [\log(1/v)] v^{d-1} \int_{-1}^1 dz (1-z^2)^{\frac{d-3}{2}}   \nonumber \\
&& \la S^{u_j u_j}(v,z) \ra,
\label{eq:tot_flux_3D_HDT}
\eea
where $A$ is given by Eq.~(\ref{eq:A}). For Eq.~(\ref{eq:tot_flux_3D_HDT}), we perform the $dz$ integral using  Gauss-Jacobi quadrature, whereas the $dv$ integral using Romberg iterative scheme. Note that the total energy flux in the inertial range is
\be
\Pi(R) = \la \Pi_{u_1} (R)\ra + (d-2) \la \Pi_{u_2} (R)\ra,
\label{eq:total_Pi}
\ee
which  equals the dissipation range $\epsilon_u$.

Now a brief discussion on the 3D energy flux, which is
\be
\la \Pi (R)\ra = \la \Pi_{u_1} (R)\ra + \la \Pi_{u_2} (R)\ra = \epsilon_u
\ee 
 Note that $\nu_{1*}$ and $\nu_{2*}$ appear in the denominators of Eqs.~(\ref{eq:Svw_u1b}, \ref{eq:SCH2(v,w)}) respectively.  Since $\nu_{1*} \ll \nu_{2*}$, the negative energy flux $\Pi_{u_1}$ dominates positive $\Pi_{u_2}$ leading to  $\Pi(R) <0$.  This is a problem! Fortunately, this issue is easily resolved by employing $\int_{0.22}^1 dv $ for $\Pi_{u_1}(R)$ of Eq.~(\ref{eq:Pi_2d}) (as in Sec.~\ref{sec:M2M_CH1}); this procedure yields $K_\mathrm{Ko} =1.64$, which is in good agreement with earlier field-theoretic computations, as well as numerical and experimental results.  As discussed in Sec.~\ref{sec:RG_u2}, $\hat{e}_1({\bf k})$ and $\hat{e}_2({\bf k})$ are transformed  to each other under the change of triads ($\hat{n}$). Therefore, we may also use $\nu_{1*} \leftarrow \nu_{2*}$ that yields $K_\mathrm{Ko}' =1.89$.

For $d$ dimension, the solution of Eq.~(\ref{eq:total_Pi})
yields $K_\mathrm{Ko}$ and  $K_\mathrm{Ko}'$ for various $d$'s. These results are listed in Table \ref{tab:nu_Ko} and illustrated in Fig.~\ref{fig:nu_Ko}.   The constant $K_\mathrm{Ko}$ increases from $d=2$ to $d=2.1$,  then decreases up to $d=4$, and finally increases again up to $d=6$.   Note that these constants are not defined for $d \ge 6$, where the equilibrium solution ($E(k) \sim k^{d-1}$ with zero flux) is valid.  Thus, $d=6$ is the critical dimension for hydrodynamic turbulence.  Also, for a given $E(k)$,   $\epsilon_u \propto K_\mathrm{Ko}^{-3/2}$. Hence, $\epsilon_u$  is inversely proportional to $K_\mathrm{Ko}^{3/2}$. 


We compare our predictions with those in the past literature.  Using Lagrangian renormalized approximation, \citet{Gotoh:PRE2007}  showed that  the Kolmogorov's constant for 3D and 4D are 1.72 and 1.31 respectively.  \citet{Berera:PF2020} reported  the corresponding constants to be  1.7 and 1.3 respectively. The corresponding numbers in our calculations, 1.63 and 1.45, are in general agreement with the earlier results. 

Exploration of turbulence in fractal dimension remains a challenge. \citet{Lanotte:PRL2015} simulated hydrodynamic turbulence in fractal dimension between 2.5 to 3. They studied variations of energy spectrum and probability distribution function of vorticity as function of fractional dimension. It will be interesting to employ similar ideas to dimension close to 2 and for much higher dimension, but we expect these numerical experiments to be very expensive.  It is possible that the fractional dimension is related to the quasi-2D anisotropic turbulence that shows a transition from positive energy flux to negative energy flux with the decrease of vertical dimension~\cite{Alexakis:PR2018}.

\section{Fractional Energy Transfers}
\label{sec:S2S_3D}

\begin{figure}
\begin{center}
	\includegraphics[scale = 0.6]{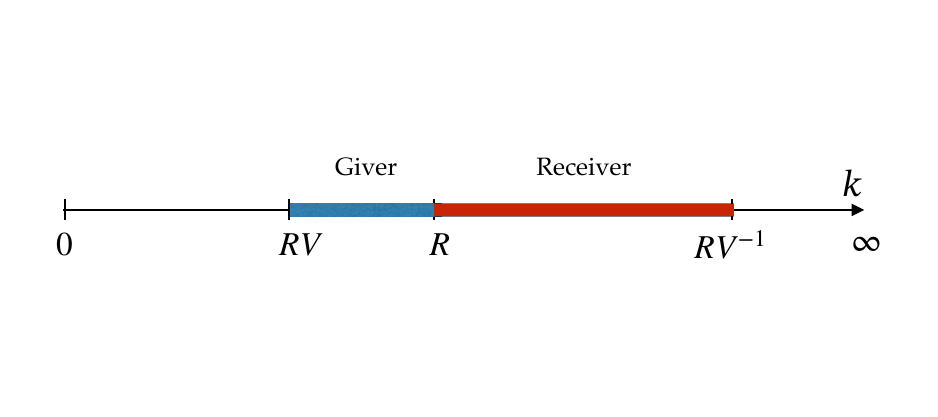}
\end{center}
\vspace*{0pt}
\caption{The fractional energy transfer $\la \Pi_{V}(R)\ra/\epsilon_u $ is  the energy transfer from wavenumber shell $(RV,R)$ to $(R, R/V)$, where $V<1$.  }
\label{fig:flux_frac}
\end{figure}

To disentangle the energy transfers between various wavenumber regimes, as well as to quantify  the dependence of the energy flux on $d$, we define \textit{fractional energy flux} as follows:
\bea
\frac{ \la \Pi_{V}(R)\ra}{\epsilon_u }  &  = & A  \int_V^1 dv   [\log(1/v)] v^{d-1} \int_{-1}^1 dz (1-z^2)^{\frac{d-3}{2}}  \nonumber \\
&&  [\la S^{u_1 u_1}(v,z) \ra + (d-1)\la S^{u_2 u_2}(v,z) \ra] \nonumber \\
& = & \int_{V}^1 dv  T(v),
\eea
where $0 < V <1$. Based on the relations of Eq.~(\ref{eq:kpq_transform}), we deduce that $\la \Pi_V(R)\ra$ represents the net energy transfer from the giver modes in the band $(R V,R)$ to the receiver modes in the band $(R, R/V)$. See Fig.~\ref{fig:flux_frac} for an illustration. 

\begin{figure}
	\begin{center}
		\includegraphics[scale = 1.0]{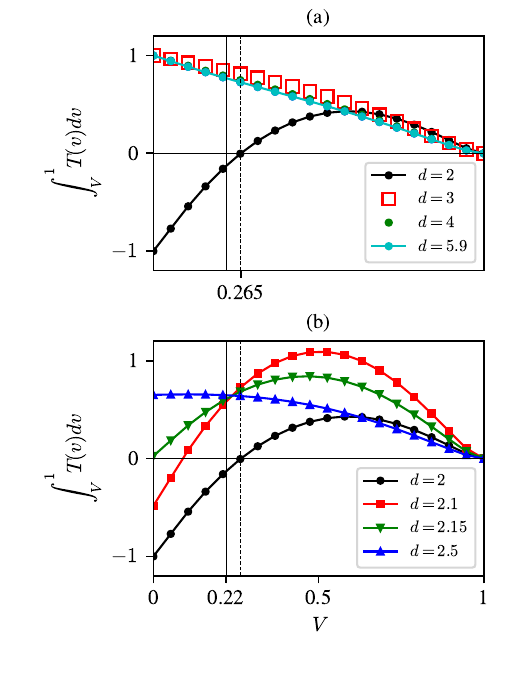}
	\end{center}
	\vspace*{0pt}
	\caption{ Fractional energy transfer $\la \Pi_{V}(R)\ra/\epsilon_u $ for various dimensions.  }
	\label{fig:Tv}
\end{figure}

We compute $\int_V^1 dv  T(v)$ for $V=(0,1)$ and $d=2,2.1,2.15,2.5, 3,4,5.9$, and plot them in Fig.~\ref{fig:Tv}(a, b).   The plots in the figure reveal that  $\Pi(R)/\epsilon_u = \int_0^1 dv  T(v)$ is negative for $d < 2.15$, positive for $d>2.15$, and  0 for $d \approx 2.15$. Hence, the energy flux $\la \Pi(R)\ra$ changes sign at $d \approx 2.15$.  The spiking in $K_\mathrm{Ko}$ near $d=2.15$ is due to the vanishing of the energy flux (see Fig.~\ref{fig:nu_Ko}).  Our results are in a reasonable agreement with those of  \citet{Fournier:PRA1978} who reported the  transition dimension for the energy flux to be approximately 2.06.

In 2D, $\int_V^1 dv  T(v) $ changes sign from negative to positive as $V$ crosses 0.265 from left to right (see Fig.~\ref{fig:Tv}(a)).  This feature arises due to the positive local transfers, but significant negative (inverse) nonlocal   energy transfers~\cite{Verma:Pramana2005S2S,	Verma:arxiv2023}. The inverse energy cascade in 2D hydrodynamics  is due to the above nonlocal reverse energy transfers. For more details on local and nonlocal energy transfers, refer to \citet{Verma:arxiv2023}.

In the next section, we compute the renormalization and Kolmogorov's constants for Kraichnan's $k^{-3/2}$ spectrum.

\section{Renormalization and energy flux computations for Kraichnan's $k^{-3/2}$ spectrum}

\citet{Kraichnan:PF1964Eulerian} argued that the sweeping effect may lead to $k^{-3/2}$ energy spectrum  for hydrodynamic turbulence.  However, experiments, numerical simulations, and analytical works  rule out this spectrum, and strongly support Kolmogorov's $k^{-5/3}$ spectrum.  Still, for mathematical curiosity we explore whether $k^{-3/2}$ spectrum satisfies the RG equation.  

\begin{figure}
	\begin{center}
		\includegraphics[scale = 0.9]{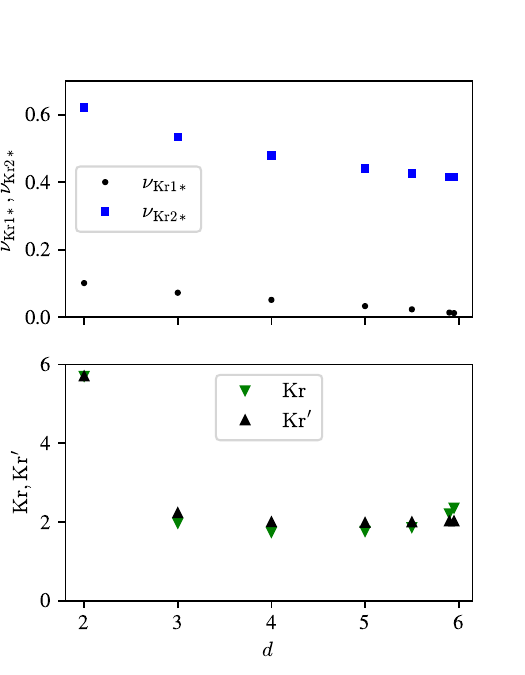}
	\end{center}
	\vspace*{0pt}
	\caption{For Kraichnan's $k^{-3/2}$ spectrum, the values of constants for various $d$'s: (a) $\nu_\mathrm{Kr1*}$ and $\nu_\mathrm{Kr2*}$; (b) $\mathrm{Kr}$ and $\mathrm{Kr}'$.  }
	\label{fig:nu_Ko_Kr}
\end{figure}

In the $k^{-3/2}$ framework,
\bea
\bar{E}(k) & = & K_\mathrm{Kr} (\epsilon_u U_0)^{1/2} k^{-3/2} ,
\label{eq:Ek_Kraichnan} \\
\bar{\nu}_{1}(k_n) & = & \nu_{\mathrm{Kr}1*} K_\mathrm{Kr}^{1/2}\ (\epsilon_u U_0)^{1/4} k_n^{-5/4}, \\
\bar{\nu}_{2}(k_n) & = & \nu_{\mathrm{Kr}2*} K_\mathrm{Kr}^{1/2}\ (\epsilon_u U_0)^{1/4} k_n^{-5/4},
\eea
where $U_0$ is the large-scale fluid velocity; $\nu_{\mathrm{Kr}1*}$ and $\nu_{\mathrm{Kr}2*}$ are  the renormalization constants for $u_1$ and $u_2$ components; and $K_\mathrm{Kr} $ is Kraichnan's constant (corresponding to Kolmogorov's constant). For the $k^{-3/2}$ energy spectrum, the Feynman diagrams and all the equations of Sections~\ref{sec:RG} and \ref{sec:ET},
except  those for $F_1(p',z)$, $F_2(p',z)$, $F_3(p',z)$, $\la {S}^{u_1 u_1}(v,z) \ra $, $\la {S}^{u_j u_j}(v,z) \ra $, $\mathrm{numr}_2$, $E(k)$, $\nu_1(k_n)$, and $\nu_2(k_n)$, are unchanged. The above equations are modified to the following form (with bar):
\bea
\bar{F}_1(p',z) & = & \frac{(1-z^2)  (p'-2z)(2p'z-1) p' q'^{-5/2-d}}{p'^{3/4} +q'^{3/4}}, \\
\bar{F}_2(p',z)  & = & \frac{(1-z^2)  (1-p'^2)(2p'z-1) p'^{-1/2-d} q'^{-2}}{p'^{3/4} +q'^{3/4}}, \nonumber \\  \\
\bar{F}_3(p',z) & = &  \frac{(1-z^2)  p'^{-1/2-d}}{\nu_{\mathrm{Kr}1*} p'^{3/4} + \nu_{\mathrm{Kr}2*} q'^{3/4}} \nonumber \\
&&+ \frac{(1-z^2) p'^2 q'^{-5/2-d}}{\nu_{\mathrm{Kr}2*} p'^{3/4} + \nu_{\mathrm{Kr}1*} q'^{3/4}}, \\
\la \bar{S}^{u_j u_j}(v,z) \ra & = &     \frac{v^2 w^{-5/2-d} (v^{-1/2-d}-1) (1-z^2)}{ \nu_{\mathrm{Kr}2*} (1+v^{3/4}) +\nu_{\mathrm{Kr}1*} w^{3/4}} \\
\la \bar{S}^{u_1 u_1}(v,z) \ra & = &     \frac{\overline{\mathrm{numr}}_2 }{ \nu_{\mathrm{Kr}1*} (1+v^{3/4} + w^{3/4})} \\
\overline{\mathrm{numr}}_2 & = &2 (2vz-1)(1-z^2) zv  w^{-2}  (vw)^{-1/2-d}   \nonumber \\
&& + 2 (v-2z)(1-z^2) zv^2  w^{-5/2-d}  \nonumber \\
&& + 2 (1-v^2)(1-z^2)  z  w^{-2} v^{-1/2-d}
\eea
for $j \ge 2 $. Using the revised equations we compute the new renormalization and Kraichnan's constants  for various space dimensions.  We observe that $\nu_{\mathrm{Kr}1*} $ has nonzero solution for $d < 6$, and it has no solution for $d>6$. However,  $\nu_{*}=0 $ is a valid solution for $d>6$.   Hence, $d =6$ is the critical dimension for the Kraichnan's spectrum as well.  In Fig.~\ref{fig:nu_Ko_Kr} we present the renormalized parameters  and Kraichnan's constant for various dimensions. Note that the constants $\mathrm{Kr} $  and $\mathrm{Kr}' $  correspond respectively to cases when $\nu_{\mathrm{Kr}1*} \ne \nu_{\mathrm{Kr}2*} $ and $\nu_{\mathrm{Kr}1*} = \nu_{\mathrm{Kr}2*} $ 

Thus, surprisingly, Kraichnan's $k^{-3/2}$ spectrum and the corresponding viscosity formulas are solutions to the RG equations for $ d < 6$. We conclude in the next section.

\section{Conclusions}

In this paper,  we employ perturbative field theory to  the incompressible Navier-Stokes equation  and compute the renormalized viscosities and  Kolmogorov's constant for various space dimensions. We employ
Craya-Herring basis  that simplifies the calculations considerably. We summarize our findings as follows.

\begin{enumerate}
	\item For space dimension less than 6,  Kolmogorov's spectrum $E(k) \sim k^{-5/3}$ is a solution of the RG equation with the renormalized viscosity scaling as $\nu_{(1,2)*} \sqrt{\mathrm{Ko}} \epsilon_u^{1/3}k^{-4/3}$, where $\nu_{(1,2)*}$ are the prefactors for the components of the Craya-Herring basis. 	These constants are computed using the recurrence relation for the renormalized viscosity. Our computed constants are in general agreement with earlier results.  These solutions are out of equlibrium when the energy flux is nonzero.
	
	\item The renormalization constants $\nu_{(1,2)*}$ are functions of space dimension. Interestingly $\nu_{1*}$  gradually decreases to zero at $d=6$.  Our detailed calculations show that the solutions for $d<6$ are out of equilibrium, but they merge with equilibrium solution at $d = 6$.   Thus, $d=6$ is the critical upper dimension. 
	
	\item For $d \ge 6$,  the viscosity remains \textit{unnormalized} ($\nu = 0$), and the equilibrium solution of Euler equation, $E(k) \sim k^{d-1}$, satisfies the RG equation. 	Note that the energy flux vanishes under this condition.  	{\citet{Adzhemyan:JPA2008} showed that  the Kolmogorov constant $K_\mathrm{Ko} \propto d^{1/3}$ which leads the vanishing energy flux, $\epsilon_u \propto K_\mathrm{Ko}^{-3/2} \propto d^{-1/2} \rightarrow 0$, as  $d \rightarrow \infty$. In similar lines, \citet{Fournier:JPA1978} showed that    intermittency vanishes as $d \rightarrow \infty$. These observations too indicate Gaussianity of the velocity field at large $d$, consistent with our results. Our   renormalization calculation indicates that the upper critical dimension for hydrodynamic turbulence is 6.  Thus, we can argue that the  nonequilibrium solution with nonzero energy flux transitions to the equilibrium solution with $\epsilon_u = 0$.  Note that the equilibrium solution with $\nu=0$ respects time-reversal symmetry. However, the nonequilibrium solution with finite $\nu$ and nonzero energy flux breaks the time-reversal symmetry~\cite{Verma:EPJB2019}. 	}


	
	\item Using field theory, we compute the mode-to-mode energy transfers, energy fluxes, and Kolmogorov's constant for $d < 6$.  The energy flux  is negative for $d <2.15$, whereas it is positive for $d>2.15$. The transition dimension $d = 2.15$  is in reasonable agreement with the predictions of   \citet{Fournier:PRA1978}, according to which the energy flux changes sign near $d \approx 2.05$.
	
	\item Our results, in particular Kolmogorov's constant, are in agreement with previous works for 4D turbulence simuations~\cite{Gotoh:PRE2007,Berera:PF2020}. Note that simulation of turbulent flows for $d \ge  4$ is very expensive due to large grid size.   Simulation of turbulence in fractional dimension is of interest~\cite{Lanotte:PRL2015}, but   these simulations too require considerable computing resources.
	
	\item { The present work does not include intermittency correction, which is more complex to compute. Researchers have employed multi-loop field-theoretic calculations~(e.g., \cite{Adzhemyan:IJMPB2003}) and Lagrangian field-theory calculations~(e.g., \cite{Falkovich:RMP2001}) to quantify intermittency in turbulence. It will be interesting to use  Craya-Herring basis  for intermittency computations. }
		
	\item Interestingly, Kraichnan's $k^{-3/2}$ energy spectrum, which is inspired by the sweeping effect, too satisfies the recursive RG equation for hydrodynamic turbulence. Note however that $k^{-3/2}$ energy spectrum is ruled out based on numerical and experimental findings, as well as from  analytical works such as Kolmogorov's K41 theory~\cite{Kolmogorov:DANS1941Structure}.

\end{enumerate}

In summary, field-theoretic tools provide valuable insights into hydrodynamic turbulence.   

\vspace{0.5cm} 
{\bf Acknowledgements:} The author thanks Soumyadeep Chatterjee for help in generating the Feynman diagrams. He also thanks Srinivas Raghu, Rodion  Stepanov, Jayant Bhattacharjee,  Krzysztof Mizerski, and anonymous referees  for comments on the paper. I got valuable suggestions on the paper during  the discussion meeting ``Field Theory and Turbulence" hosted by  International Centre for Theoretical Studies, Bengaluru.   This work is supported by  Science and Engineering Research Board, India (Grant numbers: SERB/PHY/20215225 and SERB/PHY/2021473).

\appendix
\section{Evaluation of the Renormalization and Energy Flux Integrals}
\label{sec:integration}

For the RG procedure, the integrals of  Eqs.~(\ref{eq:nu1_integral}, \ref{eq:nu2_integral}) are finite because they are performed in the band $1 \le p' \le b$ and $1 \le q' \le b$. Here, we employ the Gaussian quadrature for the $dq'$ integral and a Romberg scheme for the $dp'$ integral. This procedure yields finite and accurate results. However, the integrals for the energy flux are singular~\citep{Peskin:book:QFT} and they need special attention.

The  energy-flux integral is  of the following form:
\bea
I =  
\int_0^1 dv   [\log(1/v)] v \int_{-1}^1  dz     {(1-z^2)}^{(d-1)/2}   f(v,z),
\label{eq:Pi_2D_integral_proc}
\eea
where $ f(v,z)$ involves singularities. For accurate evaluation of $dz$ integration, we employ the Gauss-Jacobi quadrature:
\be
\int_{-1}^1  dz    f(v,z)  (1-z)^{(d-1)/2} (1+z)^{(d-1)/2}
\approx \sum_k f(v,z_k) w_k,
\ee
where $z_k$ is the $k$th root of Jacobi polynomials, and $w_k$ is the corresponding \textit{weight}. Note that   $f(v,z_k)$ is evaluated  at $z = z_k$. The Gauss-Jacobi quadrature yields finite answer for these singular integrals.  We employ a Romberg iterative scheme for the subsequent $dv$ integration.

\bibliographystyle{apsrev4-2}

%


\end{document}